\documentclass[%
reprint,
 amsmath,amssymb,
prb,
floatfix,
]{revtex4-1}

\usepackage{graphicx}
\usepackage{dcolumn}
\usepackage{bm}

\usepackage{physics}
\usepackage[mathscr]{euscript}
\usepackage[caption=false]{subfig}
\usepackage{tikz}
\usepackage[title]{appendix}
\usepackage{float}
\usepackage{hyperref}
\usepackage{lipsum, babel}
\hypersetup{
    colorlinks=true,
    linkcolor=blue,
    citecolor=blue,      
    urlcolor=blue,
}
 
\begin{document}

\title{On the signatures of non-topological patches on the surface of topological insulators}
\author{Tamoghna Barik $^1$}

\author{Jay D. Sau $^2$}

\affiliation{Condensed Matter Theory Center $^{1,2}$, Department of Physics $^{1,2}$ and Joint Quantum Institute $^2$ , \\
University of Maryland, College Park, MD-20742, USA}

\begin{abstract}
The non-trivial topology in the layered $\text{FeTe}_{0.55}\text{Se}_{0.45}$ (FTS) superconductor has been suggested by both 
theory and experiment to be strongly dependent on the Te concentration. Motivated by this together with the Te fluctuations expected from alloy disorder, we develop a simple layered model for a strong topological insulator that allows us to describe a scenario where topologically trivial domains permeate the sample. We refer to such a phase as topological domain disordered and study the local density (LDOS) of the topological surface states that can be measured using scanning tunneling spectroscopy (STS) in this phase. We find that topologically trivial domains on the surface, where one would expect the topological surface state to be absent,  appear as regions of suppressed LDOS surrounded by domain walls with enhanced LDOS. Furthermore, we show that studying the energy dependence of the STS should allow us to distinguish the topologically trivial parts of the surface from other forms of disorder. Finally, we discuss implications of such local disappearance of the topological surface states for the observation of Majorana modes in vortices.
\end{abstract}
\maketitle

\section{Introduction}
Recently an iron based chalcogenide superconductor, $\text{FeTe}_{.55}\text{Se}_{.45} $ (FTS) has been found to host a strong topological insulator (TI) phase which is both predicted by first principle calculations \cite{ZhongFang,SCZhang_TSC} and later confirmed in the experiments \cite{KanigelARPES,ShikSinARPES}. Angle resolved photo emission spectroscopy (ARPES) measurements on this FTS system have shown evidence of parity inversion at the $Z(0,0,\pi) $ point \cite{KanigelARPES} along with the existence of single Dirac dispersion spectrum at the surface \cite{KanigelARPES,ShikSinARPES,Zhang2019}.
The co-existence of superconductivity and the TI phase in the same material leads to an exciting possibility of realizing Majorana bound states (MBSs) in vortex cores~\cite{KaneMZM,hosur2011majorana,MZMinBiTe,MZMSSARinBiTe,MZMinLiFeSe}. 
Such MBSs are of particular interest as a building block for fault tolerant quantum computing \cite{KitaevQC,NayakTQCrev,Sato2017}. 
Interestingly, evidence for such MBSs in vortex cores, in the form of zero bias peaks (ZBPs) within the vortex cores have been observed 
 using scanning tunneling spectroscopy (STS) by several independent groups \cite{HanaguriZBPs,Dongfeietal,WenHanagurilike}.
 The FTS superconductors also appear to  be low density (Fermi energy) superconductors relative to the superconducting gap 
 in this system. This allows one, in principle, to separate the MBS from  Caroli-de-Gennes-Matricon (CdGM) states \cite{CdGM_FTS_2019,HanaguriZBPs} that generically exist in superconducting vortices~\cite{CdGM,CdGMstates}.
Perhaps, one of the most encouraging signatures of MBSs is the observation of nearly quantized conductance into vortex cores~\cite{zhu2020nearly}, 
which is one of the most unique aspects of MBS.

However, the ZBPs are not present in all the vortices \cite{HanaguriZBPs,WenHanagurilike} and the conductance plateau for most of the ZBPs are significantly less than unity \cite{zhu2020nearly}. 
The reduction of the percentage of ZBPs within vortices with increasing magnetic field seen in \cite{HanaguriZBPs,WenHanagurilike} has been argued to arise due to increased coupling between nearby MBSs with decreasing inter-vortex distances in a recent theoretical work \cite{Majoranacouplingtheory}. Another proposed explanation for the disappearance of MBSs is a topological phase transition in line vortex driven by fluctuations of iron impurity concentration \cite{GhaemiZeemancoupltheory}. While the former explanation will be relatively benign to MBSs at low density, the latter mechanism would be associated with quasiparticle poisoning that would be detrimental to a Majorana qubit. Such quasiparticle poisoning is consistent with the suppression of the conductance 
height for most of the ZBPs seen in recent experiments \cite{zhu2020nearly}. However, a more complete understanding of disorder that can affect the details of vortex MBSs is still lacking in the literature. One example that we focus on in this work is the role of Te/Se composition fluctuations on the topological surface states that MBSs rely on.

The importance of the Se/Te concentration fluctuations becomes obvious on considering the fact that Te doping is necessary to drive topologically trivial FeSe into a non-trivial phase, $\text{FeTe}_{1-x}\text{Se}_x $ (FTS) with $x=0.45$. The dependence of the topological character in FTS systems on $x$ is further supported by first principle calculation \cite{ZhongFang} which finds that increasing Te concentration, ($1-x $), shifts the center  of the $p$-type band to lower energies facilitating band inversion. This sensitivity of the topological nature of FTS to Te doping is consistent with recent ARPES experiments \cite{Brookhaven} where the topological surface states are found to disappear below a certain Te doping. Since in $\text{FeTe}_{.55}\text{Se}_{.45} $ the topological phase appears to occur in the alloy phase, fluctuations in $x$ are likely to occur in much of the sample. This can lead to local variations of the topological invariant on the surface in which case the topological surface states may disappear from parts of the surface. As we will discuss in the next section, we expect small fluctuations in $x$ to be able to drive such variations in the topological invariant. We will refer to this phase as a topological domain disordered phase. As we will discuss in Sec. \ref{sec:conclusion}, such local fluctuations that lead to the local disappearance of the surface state from the top layer of the sample are expected to affect the properties of MBSs in the system as probed by STS.

While variation of parameters in the Hamiltonian of any strong TI can lead, in principle, to the fluctuation of a local topological invariant \cite{hastings2011topological,Estienne2012}, the layered nature of FTS provides a system where the topological invariant can vary on the scale of a single atomic plane. This, potentially, allows access to such domain wall physics using STS. From a theoretical stand-point, this layered nature allows us to construct a relatively simple phenomenological model of the topological phase by considering nearest neighbor tunneling between the layers of electrons. 
The phenomenological model is determined almost entirely by the symmetries (i.e. properties under $C_4$ rotation) of the low-energy bands near zero momentum  and is independent of the complex details of the strongly correlated spectrum of a layered material. In this work, we choose parameters so that the bulk bands and the computed surface state dispersion match the surface spectrum measured in ARPES  \cite{KanigelARPES,ShikSinARPES} within the relevant energy and momentum scale (Sec. \ref{sec:Idealsurfstate}). We then use this model to study the effect of a disorder potential in the band-inversion parameter on the LDOS within the energy window of Dirac dispersion (Sec. \ref{sec:Disordsurfstate}).  This disorder potential realizes the topological domain disordered phase. Finaly, we contrast the variation of surface LDOS to the more conventional case of chemical potential disorder to provide a signature of the topological domain disordered phase that is detectable in experiment (Sec. \ref{sec:topovschem}).

\section{The effective model}\label{sec:EffMod}
\subsection{Motivation}\label{subsec:Motivation}
\begin{figure}[ht]
	\centering
    \includegraphics[width=\columnwidth]{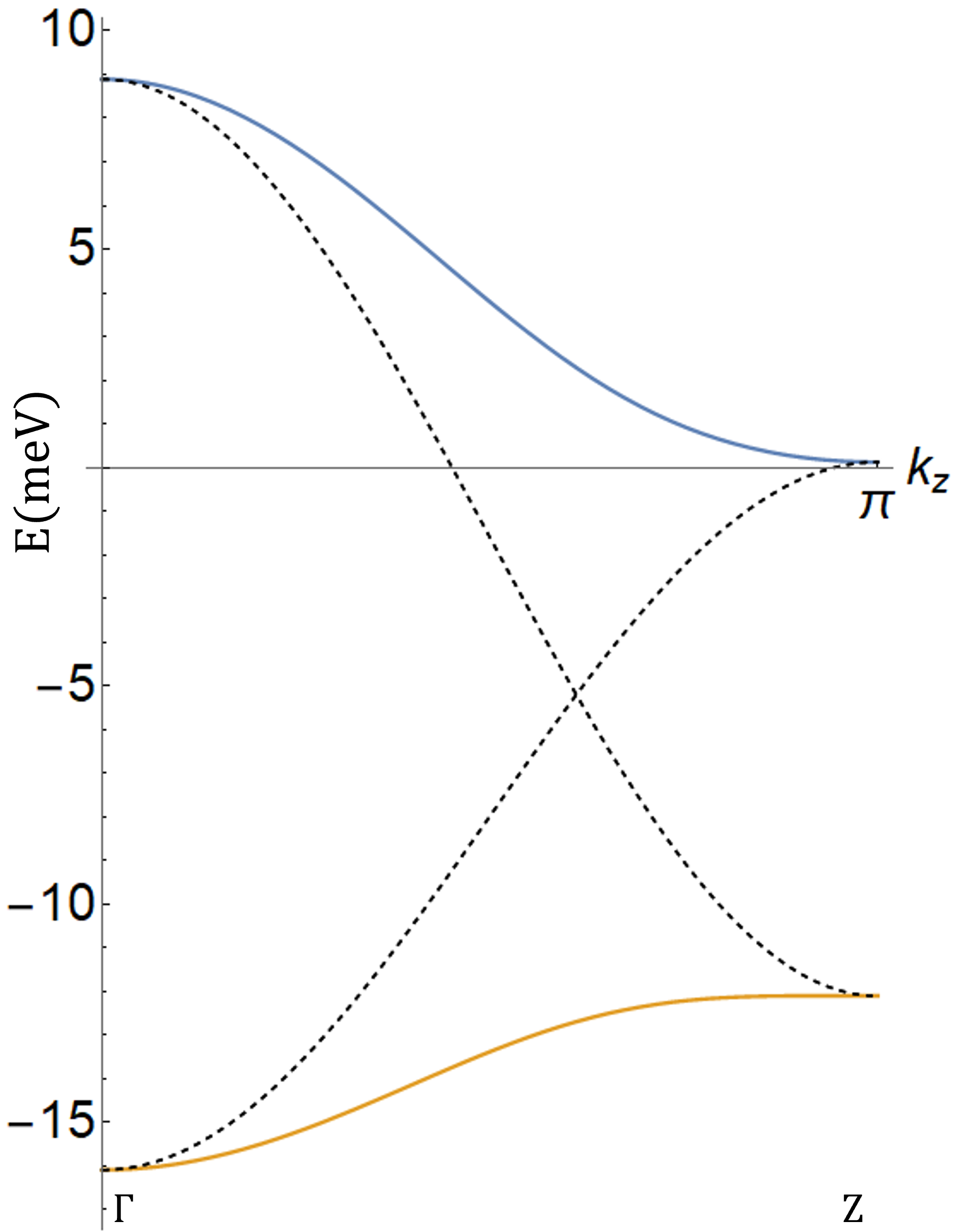}\label{fig:onlybandinversion}
    \caption{$\Gamma Z $ band dispersion without SOC (dashed curve) and with SOC (solid curve) are shown. Crossing between d-type and p-type bands happens for $|\delta _{\parallel}|<|\delta _z| $ (dashed). Hybridization between them via SOC opens up a gap (solid).}
    \label{fig:bandinversion}    
\end{figure}

In order to gain insight required to develop an effective model for the topological surface states of FTS, 
let us consider  the dispersion for the FTS system along the $\Gamma Z $ line (Fig. \ref{fig:bandinversion}), which shows a band-inversion between the even parity $d_{xy}$-band and the odd parity $p_z$-band \cite{ZhongFang}. This band-inversion, absent in FeSe (i.e. $x=1$), along with an SOC induced insulating gap is responsible for the topological surface states.   
The top of the valence band at the $Z $ point has inverted parity like $p $-band  in Fig. \ref{fig:bandinversion} is seen 
to be only about $20 \, \text{meV}$ below the Fermi energy as found in experiments \cite{KanigelARPES}. This 
value is significantly smaller compared to first-principle calculations \cite{ZhongFang} where the $p$- band is found to be about $0.5 \, \text{eV}$ below the Fermi energy at $Z $. 
Assuming that the large shift $\sim 1.5 eV$ of the $p$- band with $\text{Te}$ concentration changing from $(1-x)=0$ to $(1-x)=0.5$, 
as found in first principles calculations~\cite{ZhongFang}, is qualitatively consistent with the real material, one would expect the composition $x=0.45$ to be precariously close 
to the topological phase boundary by only about $\delta E\sim 0.5*0.02/1.5 \, \text{eV}\sim 5\, \text{meV}$.

The fact that the composition of $\text{FeTe}_{.55}\text{Se}_{.45} $ is near the topological phase boundary appears to be consistent with a recent spatially resolved ARPES experiment \cite{Brookhaven}.
Spatially resolved electron diffraction spectroscopy (EDS) analysis of a large sample revealed significant fluctuations in the $\text{Fe/Te/Se }$ density distribution over various length-scales.
This is consistent with atomic scale topographic image of the surface of $\text{FeTe}_{.55}\text{Se}_{.45} $ \cite{Vidyadensityinhomogeneity} which shows spatial in-homogeneity in $\text{Te} $ density profile. 
 The ARPES experiment in \cite{Brookhaven} correlated the spatially resolved ARPES data to EDS data to estimate the composition dependence of the phase diagram.
Using this analysis, the experiment \cite{Brookhaven} has found that 
$\text{FeTe}_{.55}\text{Se}_{.45} $ though topological resides close to the phase boundary between the topological and trivial phases since small variations of the Te concentration are seen to destroy the topological surface states.
\subsection{Model Hamiltonian\label{subsec:ModHam}}
We construct an effective model for the strong TI phase including a pair of bands with opposite parity ($d$-type and $p $-type). 
 We label the even ($d$-type) and odd ($p $-type) parity bands by $\rho_z= \pm 1 $ respectively. Adding the $j_z = \pm 1/2 $ angular momentum degree of freedom described by the set of Pauli matrices $\sigma_{\alpha=x,y,z}$, leads to a model similar to the Bernevig-Hughes-Zhang (BHZ)
model for each layer \cite{BHZmodel} as elaborated below.
Our model for FTS respects parity described by the operator $\rho_z$ and TRS described by the operator $i \sigma ^y K$. Here $K$ is the complex conjugation operator. The four fold rotational symmetry around the $z $-axis ($C_4 $) for FTS results in an effective rotational symmetry at long wave-lengths. The $p_z$-type band that we consider is invariant under $C_4$ rotation and has orbital angular momentum $l_z=0$ and hence total angular momentum $ j_z=\pm 1/2$ including spin.
In contrast, the $d_{xz}\pm i d_{yz}$ bands have orbital angular momenta $l_z=\pm 1$. A large spin-orbit splitting gives rise to a pair of d bands with total angular momenta $j=l+s= 3/2, \, 1/2$. 
Thus, combining the two orbital sectors, we obtain two sectors with total angular momentum $j_z= \pm 1/2 $ (one of $d_{xy} $-type and one $ p_z$-type) but with opposite parity $\rho_z = \pm 1 $.
By applying the appropriate unitary transformation, we can define $\sigma_z=2j_z$ in this manifold.
This implies the $C_4 $ rotation operator to be $C_4 = e^{i \pi j_z/2}=e^{i \pi \sigma ^z/4} $. 
In the more rigorous group theory representation (e.g. in the DFT calculation \cite{ZhongFang}), the states $\rho_z=\pm 1$ correspond to states in the $\Gamma_{6}^{\pm}$ representations at the $\Gamma$ and $Z$.
 Focusing near small in-plane momenta  (i.e. $k_x, k_y \simeq 0 $), the Hamiltonian in this space of $\Gamma_{6}^{\pm}$ states,
 under the symmetry constraints and the approximation of nearest neighbor tight-binding along the $z$ direction can be written as:
\begin{align}\label{eq:bandinversionham}
	\begin{split}
		&\hat{H}_I(k_{\parallel} \simeq 0, k_z) \\
        &= \left[ \delta _{\parallel}+ \delta _z \, \cos k_z -\frac{A}{2}\left( k_x^2+k_y^2 \right)\right] \, \sigma ^0 \, \rho ^z + t' \, \sin k_z \, \sigma ^z \, \rho ^x.
	\end{split}
\end{align}
Here the parameters $\delta_{\parallel,z}$ determine the energy differences between the $\Gamma_{6}^{\pm}$ states at the $\Gamma$ and $Z$ point, which are written as $\delta_{\parallel}+\delta_z$ and $\delta_{\parallel}-\delta_z$ respectively. Thus, if $|\delta _z|>|\delta _{\parallel}| $ the two bands can cross somewhere along the $\Gamma Z $ line facilitating band inversion (see Fig. \ref{fig:bandinversion}). 
In our model (Eq. \ref{eq:bandinversionham}) it corresponds to the coefficient of $\rho ^z $ changing sign while going from $\Gamma$ to $Z$, which implies the valence band states have opposite parity eigenvalues at these two TRIMs. This is the condition for a band-inversion to occur which leads to a non-trivial strong TI phase \cite{KaneZ2invariant}. The second term describes spin orbit coupling (SOC) which hybridizes the two d and p type $\rho_z = \pm 1/2 $ sectors to open up a gap along $\Gamma Z  \, (k_z : 0 \to \pi )$ direction (Fig. \ref{fig:bandinversion}), which is required for the system to have an insulating gap.  
Note that while FTS is a superconducting material, a topological surface state in a range of in-plane momentum $k_{x,y}$ can only exist in a range of energy where there is a gap as the perpendicular momentum, $k_z $, changes. This does not
preclude a superconducting state from gapless states at different in-plane momenta from the topological surface states. 
The flatness of the valence band along $\Gamma Z $ (band width of $\leq 4 \text{meV} $ as in \cite{KanigelARPES}) justifies the tight binding approach used to model the dispersion along $k_z $ direction. The parameter $A$ in the first term of $\hat{H}_I(k_{\parallel} \simeq 0, k_z)$ determines the curvature of the parabolic dispersion for small in-plane momenta ($k_{\parallel} \simeq 0 $) around the $\Gamma $ point.  A system with the Hamiltonian Eq. \ref{eq:bandinversionham} with $\delta_{\parallel,z}$ chosen to be in the topological regime ($|\delta _z|>|\delta _{\parallel}| $) hosts topological surface states on the $(001 )$ surface with in-plane Dirac dispersion. The corresponding velocity for the dispersion near the $\Gamma $ point is determined by a SOC term given by,
\begin{align}\label{eq:SOCham}
	\begin{split}
		\hat{H}_{SOC} = \alpha \left( k_x \, \sigma ^x + k_y \, \sigma ^y \right) \, \rho ^x
	\end{split}
\end{align}
Finally, since the ARPES spectrum of the FTS system doesn't appear to show particle-hole symmetry \cite{KanigelARPES,ShikSinARPES}, we include the corresponding symmetry breaking term as follows
\begin{align}\label{eq:ehham}
	\begin{split}
		& \hat{H}_{eh} = \delta _2 \, \cos k_z \, \sigma ^0 \, \rho ^0
	\end{split}
\end{align}
The three terms described above can be combined to obtain our model Hamiltonian
\begin{align}\label{eq:fullmodelham}
	\begin{split}
		& \hat{H} = \hat{H}_I + \hat{H}_{SOC}+\hat{H}_{eh}
	\end{split}
\end{align} 
which can describe the surface states of FTS in the topological phase with topological Dirac surface states along with a bulk spectrum that is consistent with ARPES measurements of the $(001) $ surface  for parameters described later. 

\begin{figure}[ht]
	\centering
    \includegraphics[width=\columnwidth]{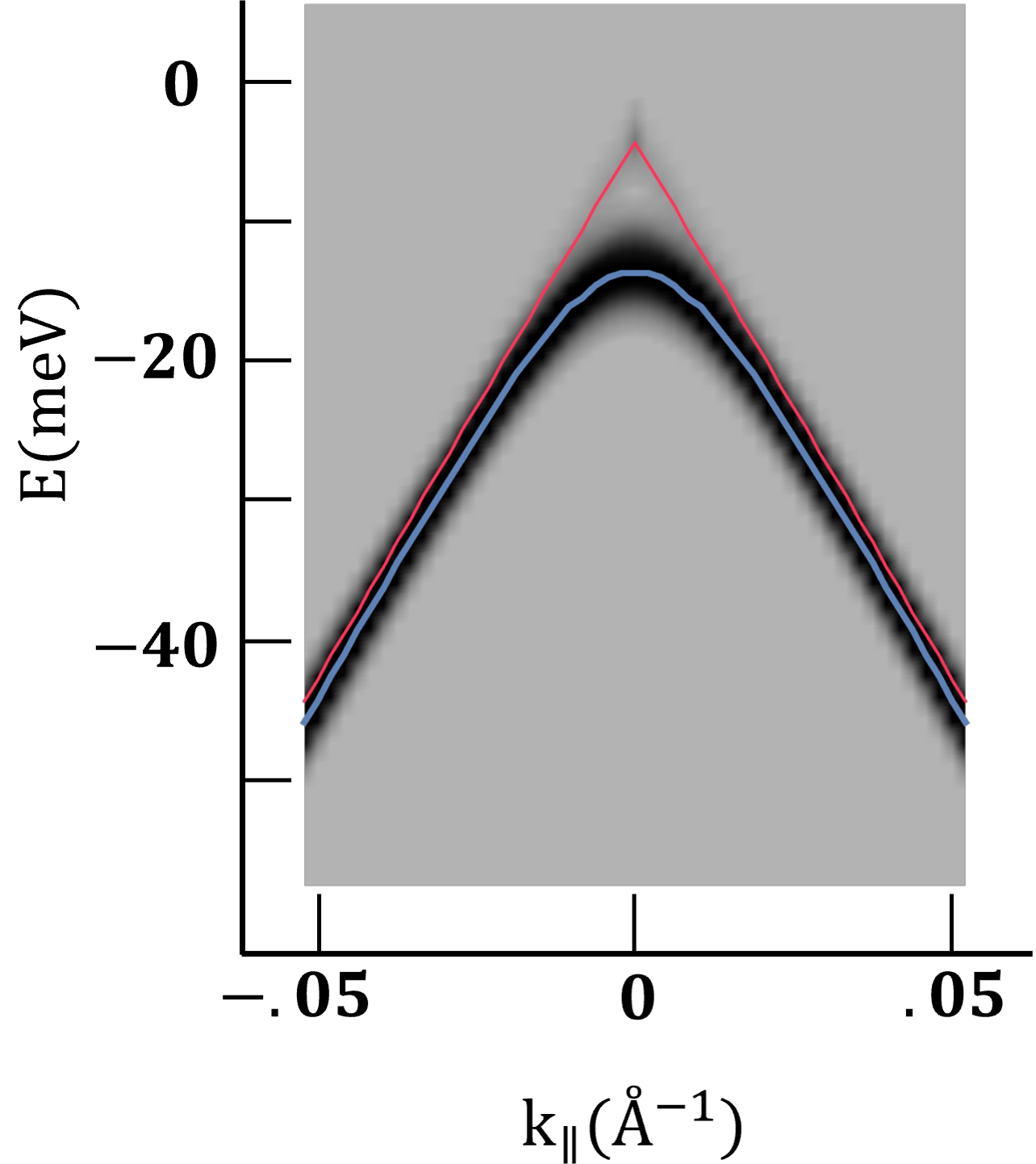}
    \caption{Surface-weighted parabolic bulk dispersion and the Dirac dispersion for the surface states along with their energy dispersion curves (EDC) shown in blue and red respectively.}
    \label{fig:inplanedispersion}    
\end{figure}

\section{Numerical results for surface state spectra}
\subsection{Ideal surface state \label{sec:Idealsurfstate}}
For the purpose of simulating surface states, it is necessary to represent the Hamiltonian in Eq. \ref{eq:fullmodelham} in real space along the $z$-direction. We do this by discretizing the z-direction with a lattice parameter c. The resultant Fourier transform translates $\cos{k_z}$ to $\cos{k_z}\rightarrow (|z\rangle\langle z+c|+h.c)/2$ and $\sin{k_z}$ by $\sin{k_z}\rightarrow i(|z\rangle\langle z+c|-h.c)/2$. Here $z$ is the real-space coordinate representing planes stacked in the $z$ direction. Using the resulting discretized Hamiltonian we simulate the surface states for each $(k_x, \, k_y) $ around the $\Gamma  $ point numerically.
 
The calculated surface dispersion and the bulk bands near the $\Gamma $ point are shown in Fig. \ref{fig:inplanedispersion}. The energy dispersion curve (EDC) for the broad parabolic bulk band has been shown by the blue line and that of the Dirac dispersion by the red line. The parameters in Eqs.  \ref{eq:bandinversionham} $\mbox{-} $ \ref{eq:ehham} are tuned such that the relevant quantities from the simulated dispersion match the estimates from the ARPES measurements in \cite{KanigelARPES,ShikSinARPES}, viz. (i) the band width of the valence band along $\Gamma Z $ direction in Fig. \ref{fig:bandinversion} (about $4 \, \text{meV} $ \cite{KanigelARPES}), (ii) the energy gap between the Dirac point and the top of the bulk valence band ($\sim 10 \, \text{meV}$) and (iii) the Dirac velocity for the surface states dispersion ( $\sim 370 \, \text{meV-{\AA} }$) , the latter two of which are measured in Ref. \cite{KanigelARPES,ShikSinARPES}. In ref. \cite{KanigelARPES}, there are similar dispersion for the surface states from a phenomenological model like us but with important differences in the quantities described above.

\subsection{Disordered surface state \label{sec:Disordsurfstate}}
We now use the parameters determined in the previous paragraph to study tunneling properties of the topological domain disordered phase. As discussed in the introduction, such a phase can arise from local fluctuations in the Se concentration $x$. We model these fluctuations by allowing the parameter $\delta _{\parallel}$ in the Hamiltonian (Eq. \ref{eq:bandinversionham}) to vary in space ($x \mbox{-}y\mbox{-}z$) according to the relation  $\delta _{\parallel}(x,y,z) = \delta _{\parallel}+w(x,y,z) $ where $w(x,y,z) $ is random with a Gaussian distribution. Motivated by the layered structure of FTS, we assume $w(x,y,z)$ to be smooth in each $x\mbox{-}y$ plane, but uncorrelated between neighboring layers.
For the purpose of conceptual and also computational simplicity, we start by assuming that $w(x,y,z)$ varies along the $x$ and $z$ direction but is uniform along the $y$-direction i.e. $w(x,y,z)\equiv w_{1D}(x,z)$. As mentioned earlier, we assume that the disorder fluctuations are uncorrelated between different layers along $z$. However, motivated by the strong in-plane electronic dispersion \cite{ShikSinARPES} as well as correlation between the positions of Se atoms in the plane~\cite{Vidyadensityinhomogeneity,HanaguriZBPs}, we assume that $w_{1D}(x,z)$ has a finite correlation length along the $x$ direction that is represented by it's Fourier transform along $x$ having the form:
 \begin{align}\label{eq:disorderdefn}
 	\begin{split}
 		&w_{1D}(k_x,z) = \frac{\sigma _{w}}{\sqrt{2}} \, \exp \left[ -\frac{\left( k_x \lambda \right)^2}{2} \right] \, \left[ X+i \, Y\right]. 
 	\end{split}
 \end{align}
In the above, $X $ and $Y$ are two random variables for each value of $k_x,z$, which are chosen from the normal distribution, $N(0,1)$. $\sigma _{w} $ and $\lambda $ are respectively the amplitude and characteristic length scale of the disorder potential. 
 We estimate the typical length scale of Te density variation to be $3 \, \text{nm} $ from STS topographic images from the experiments \cite{Vidyadensityinhomogeneity}  and use that as the value for $\lambda $. 
Since the potential $w_{1D}(x,z)$ breaks translation invariance along $x$, it is necessary to replace the $k_x$ momentum in Eq.~\ref{eq:bandinversionham} by $k_x\rightarrow -i\partial_x$, 
where $\partial_x$ is a discretized derivative with an in-plane lattice parameter $a$. 

\begin{figure}[ht]
	\centering
    \includegraphics[width=\columnwidth]{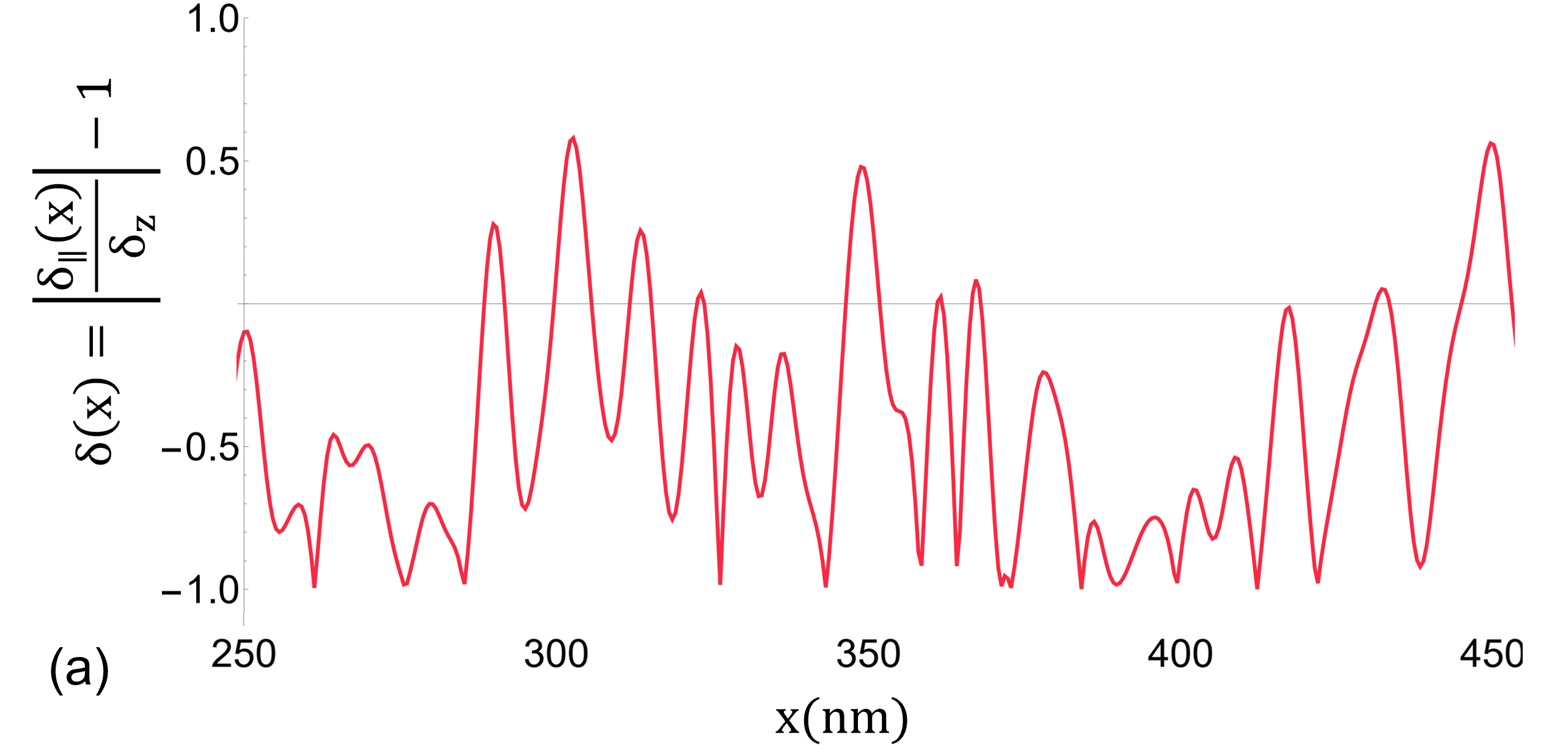}\label{fig:topodisorder_pot}
    \\
    \includegraphics[width=\columnwidth]{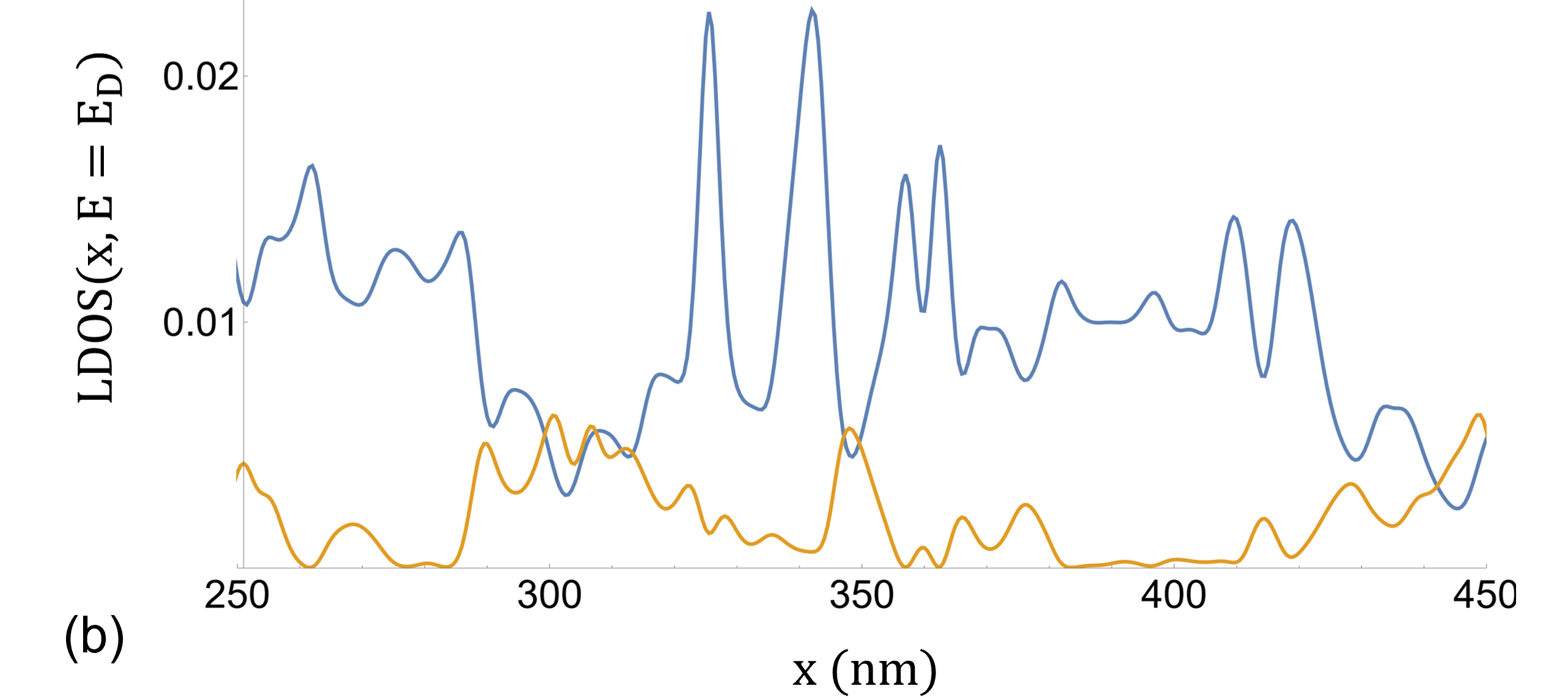}\label{fig:topodisorder_LDOS}
    \\
  \includegraphics[width=\columnwidth]{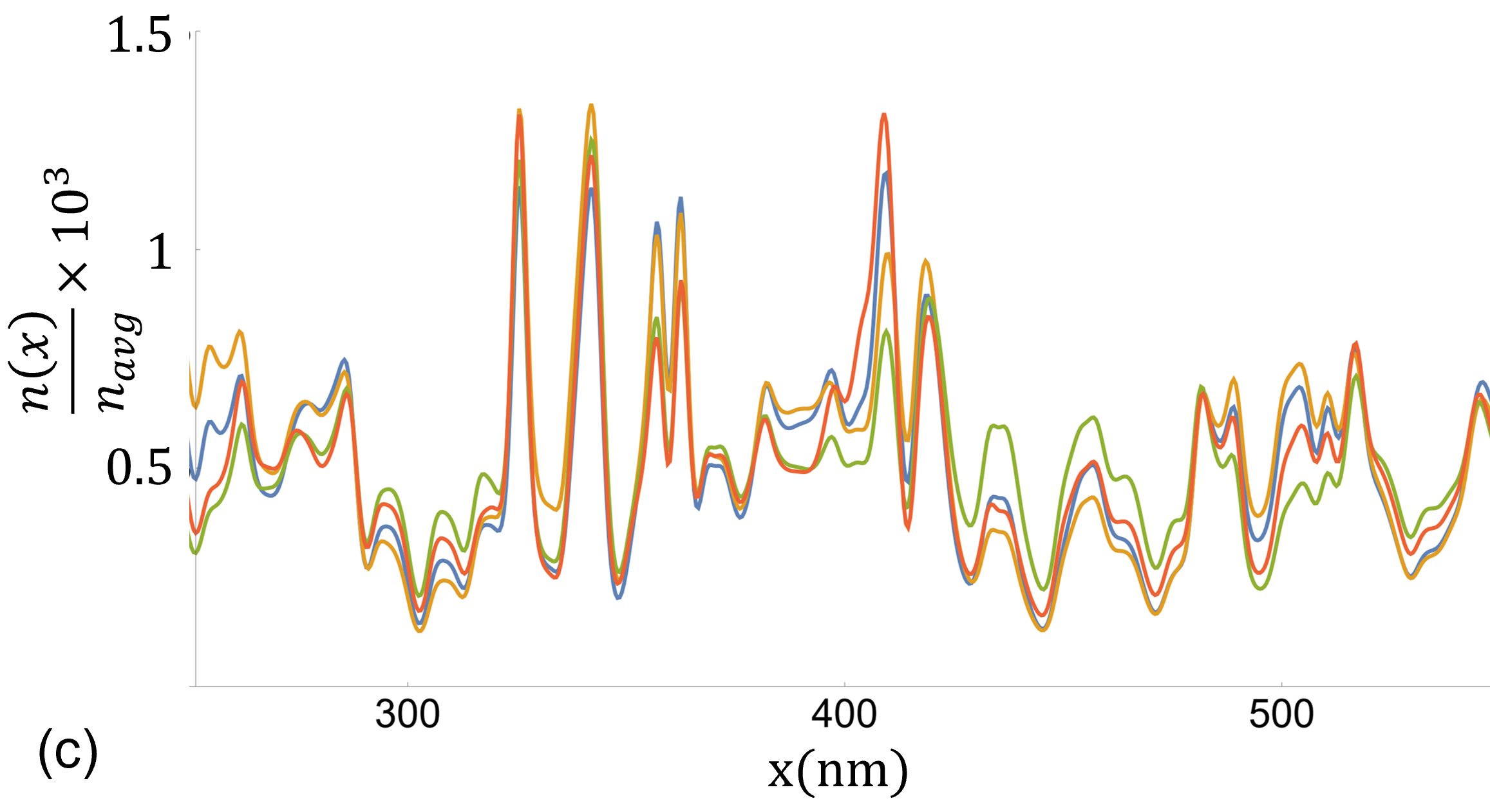}\label{fig:topodisorder_LDOS_scaling}
    \caption{(a) $|\delta _{\parallel}(x)|/|\delta _z|-1 $ plotted as the red curve where the black lines show some of the topological ($\delta (x)<0 $) and trivial ($\delta (x)>0 $) phase boundaries. (b) Top layer (blue) and second layer (yellow) LDOS for the topological domain disorder. (c) Normalized top layer LDOS at different energies within the Dirac dispersion window collapse on each other.}\label{fig:oneD_topodisorder}
\end{figure}

The numerical results for the surface density of states in the case where $\delta_{\parallel}(x,y,z)$ varies along the $x$ and $z$ directions, which is  calculated using the approach described in the previous paragraph are shown in Fig. \ref{fig:oneD_topodisorder}. Here Fig. \ref{fig:oneD_topodisorder}(a) shows the profile of $\delta_{\parallel}(x)$ on the top layer i.e. $z=0$.
 As discussed in Sec. \ref{subsec:ModHam}, regions with $|\delta _{\parallel}|< | \delta_{z}|$ are in a nominally trivial phase, while the rest of the surface is in the topological phase. To help identify trivial regions, Fig. \ref{fig:oneD_topodisorder}(a) plots $\delta(x)=|\delta_{\parallel}/\delta_z|-1$, so that trivial domains on the otherwise topological surface would appear as regions with $\delta(x)>0$. As discussed in the introduction, this local attribution of trivial and topological is more as a guide to understanding the numerical results that follow rather than a strict identification (since a topological phase is strictly speaking a global property).
The LDOS associated with this disorder realization applied to the Hamiltonian Eq. \ref{eq:fullmodelham} is plotted for the top (i.e. $z=0$) and the second layer (i.e. $z=c$) of the system in Fig. \ref{fig:oneD_topodisorder}(b). We notice that  the top layer LDOS is suppressed at the nominally trivial regions whereas the second layer LDOS is enhanced. The same observation holds for different energies within the Dirac dispersion window. In fact, the normalized LDOS for the top layer at different energies nearly collapse on each other as shown in Fig. \ref{fig:oneD_topodisorder}(c) if scaled appropriately. This suggests that the conclusion that  the Dirac surface states are transferred from the top layer to the layer below in the locally trivial region applies over a range of energy. This transfer of LDOS between the layers would be difficult to measure in STS. However, a closer examination of Fig. \ref{fig:oneD_topodisorder}(b) shows that the trivial regions represented by LDOS enhancement in the second layer occur in the vicinity of large LDOS peaks in the top layer near the topological-trivial boundaries. Such peaks in the LDOS on the top layer that demarcate the trivial domains may be visible in STS.

\begin{figure}[ht]
	\centering
  \includegraphics[width=\columnwidth]{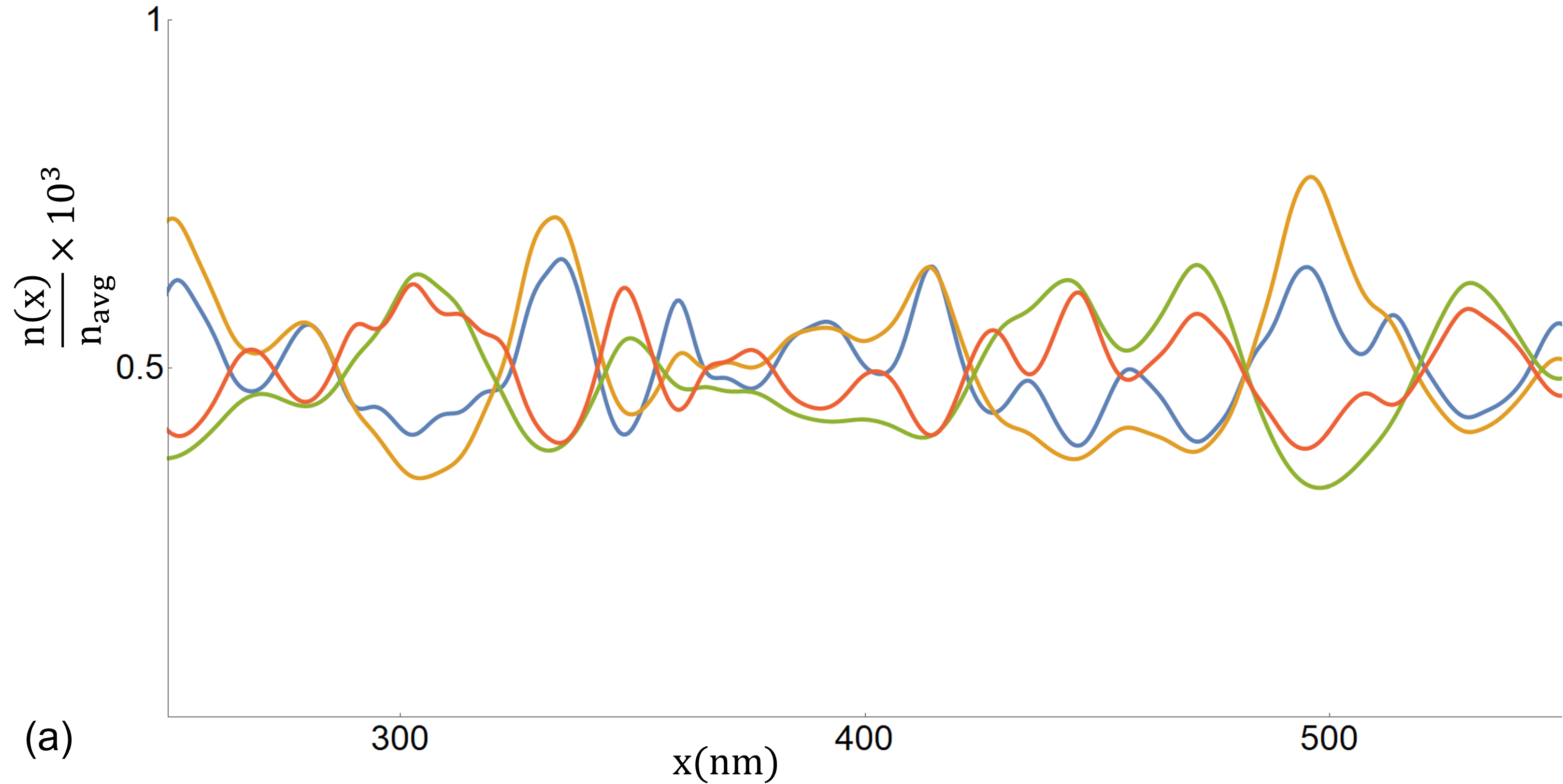}\label{fig:chemdisorder_LDOS_scaling}
    \\
    \includegraphics[width=\columnwidth]{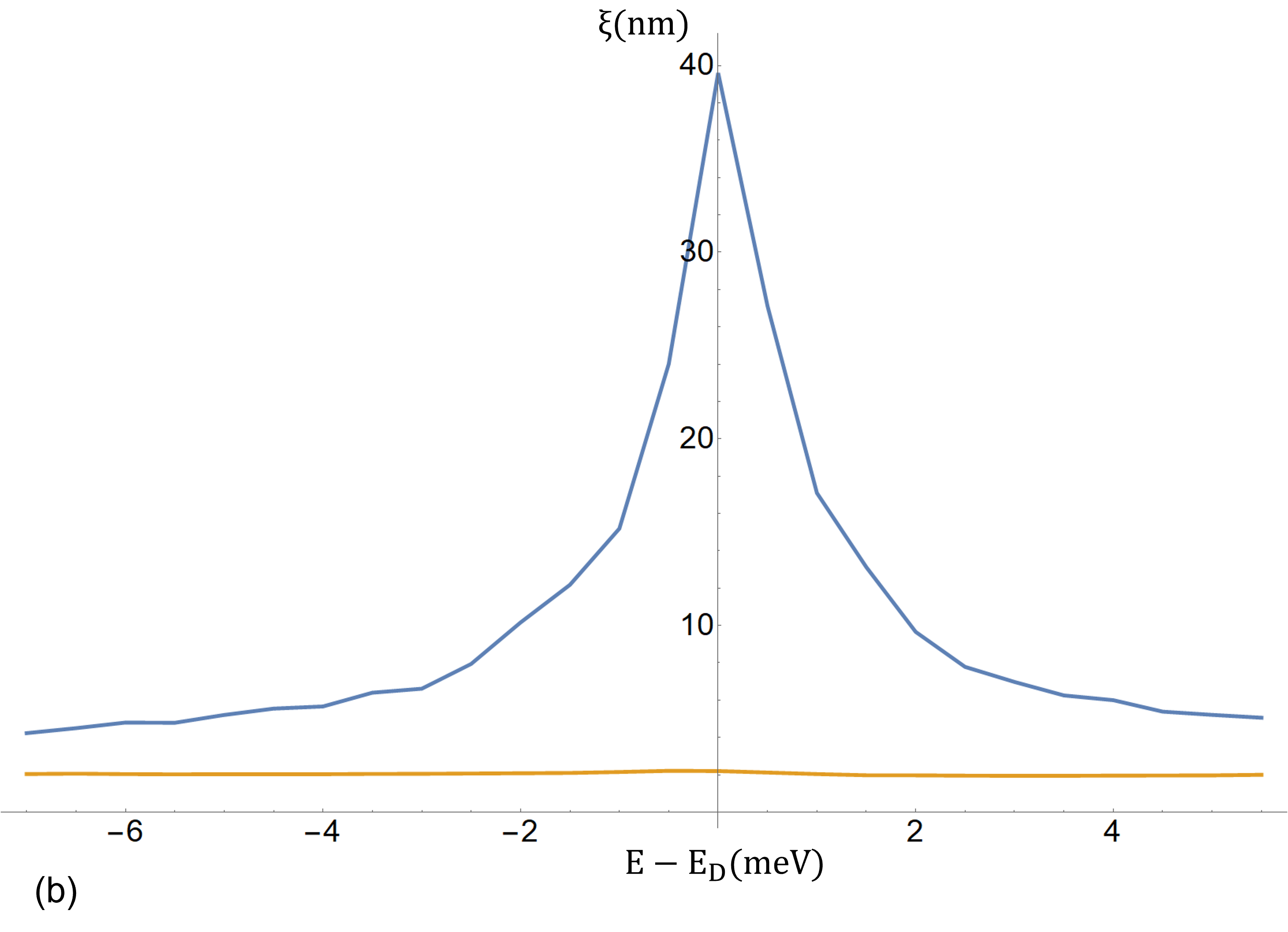}\label{fig:topovschem_LDOS_xi}
    \caption{(a) Normalized LDOS at different energies for the chemical potential disorder. 
(b) Comparison of energy dependence of LDOS-LDOS correlation length between the topological domain disorder (orange) and the chemical potential disorder (blue) cases}
\label{fig:topovschem}
\end{figure}

\subsection{Topological domain disorder versus chemical potential disorder \label{sec:topovschem}}
The more obvious signature of a trivial patch on the surface of a strong topological insulator would be the reduced local density of states seen in Fig. \ref{fig:oneD_topodisorder}(b). However, as also apparent from this plot, dips in local density of states are also seen in areas of strong fluctuations of $\delta(x)$, which are otherwise topological. This is consistent with the fact that disorder-induced Fermi energy fluctuations on the surface can lead to the Dirac point crossing the local Fermi level. The local density of states would be expected to be suppressed at such a point, since the density of states of the Dirac dispersion in two dimensions vanishes at the Dirac point.
 To determine distinctive features associated with topological domain disorder relative to conventional charge density fluctuations, we simulate the LDOS in the presence of chemical potential disorder with a Gaussian distribution. For this purpose, we keep $\delta_{\parallel} $ fixed and add a Gaussian disorder potential proportional to the identity matrix, $w (x) \sigma ^0 \, \rho ^0 $, to the Hamiltonian in Eq. \ref{eq:fullmodelham}. We choose the amplitude of chemical potential fluctuations to be about $2 \, \text{meV}$ (estimated from the broadening of the Dirac point in ARPES measurement \cite{KanigelARPES,ShikSinARPES}) and choose the same length scale ($\lambda \simeq 3 \text{nm} $) for the disorder variation similar to Eq. \ref{eq:disorderdefn}. Looking at the normalized top layer LDOS plot at several energies (Fig. \ref{fig:topovschem}(a)), we notice that, unlike the case of topological domain disorder, the LDOS profiles at various energies are distinct in shape and do not collapse on each other upon scaling in stark contrast to the case of topological domain disorder. The difference of the profiles in this case for different energies is related to the dependence of the  length scale for LDOS variation, $\xi$, with energy as $\xi (E)\propto 1/(E-E_{D}) $ around the Dirac energy $E_D $ (Fig. \ref{fig:topovschem}(b)).\\
Hence, the LDOS variation within the Dirac dispersion window are distinct between the cases of topological domain disorder and chemical potential disorder. The two cases can be distinguished by comparing the normalized LDOS at various energies and the energy dependence of the characteristic length scale for LDOS variation, $\xi (E) $ for the top layer. Thus, the LDOS peaks in Fig.~\ref{fig:oneD_topodisorder}(b) that are in the vicinity of a topologically trivial patch, can potentially be distinguished from peaks associated with Fermi energy fluctuations based on their energy independence.
Besides the energy independent peaks, another signature of the LDOS features associated with topological domain disorder is the energy independence of the characteristic length scale. We estimate the characteristic length scale $\xi$ for the LDOS variation as the inverse of the width, $\Delta k_x$, of the Fourier spectrum of the LDOS (i.e. quasiparticle interference spectrum) according to the relation $\xi \simeq 1/\sqrt{\langle \Delta k_x^2 \rangle} $.
 Fig. \ref{fig:topovschem}(b) shows a plot of the length-scale $\xi$ as a function of the tunneling energy relative to the Dirac point.  We find that for chemical potential disorder (shown in blue) $\xi$ increases substantially as energy is reduced. This is in contrast to topological domain disorder (shown in yellow) where the associated length-scale $\xi$ is seen to be energy independent in Fig. \ref{fig:topovschem}(b).  This conclusion is consistent with the collapse of the various LDOS peaks seen in Fig.~\ref{fig:oneD_topodisorder}(c), which shows that the widths of peaks at the edge of trivial regions are independent of energy. This can be understood as these peaks being associated with tunneling into domain walls, whose widths are controlled by spatial variation of the disorder parameter $\delta(x)$.\\
\begin{figure}[ht]
	\centering
    \includegraphics[width=\columnwidth]{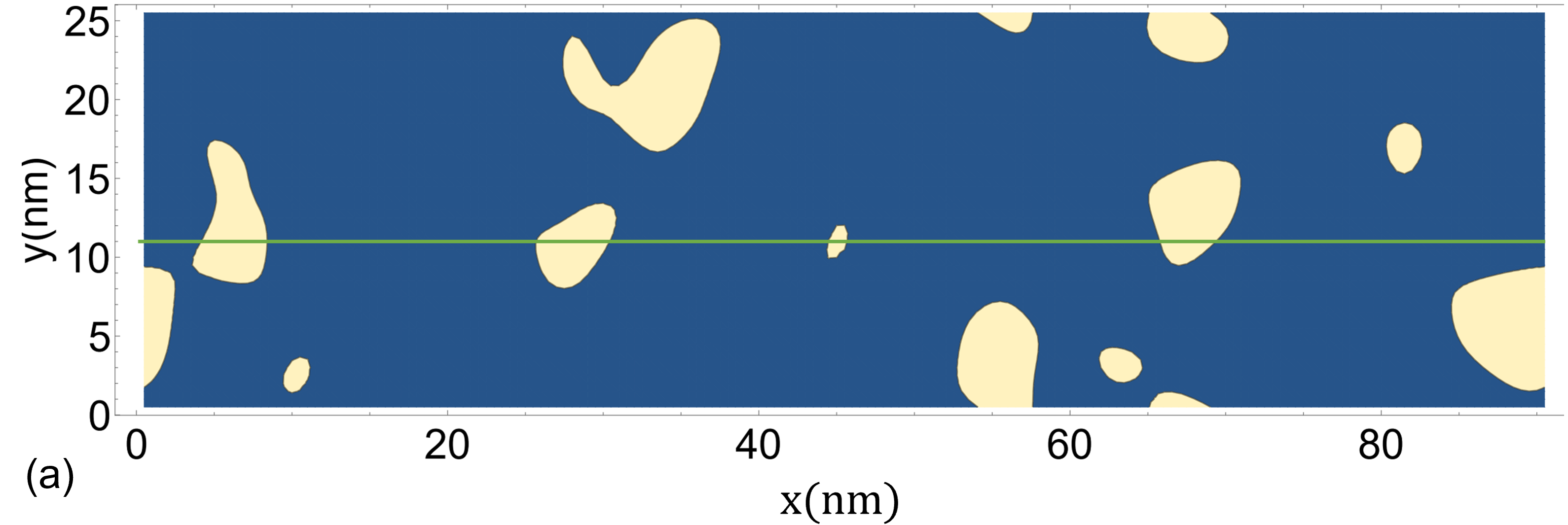}\label{fig:2D_LDOS_top_pot}\\
    \includegraphics[width=\columnwidth]{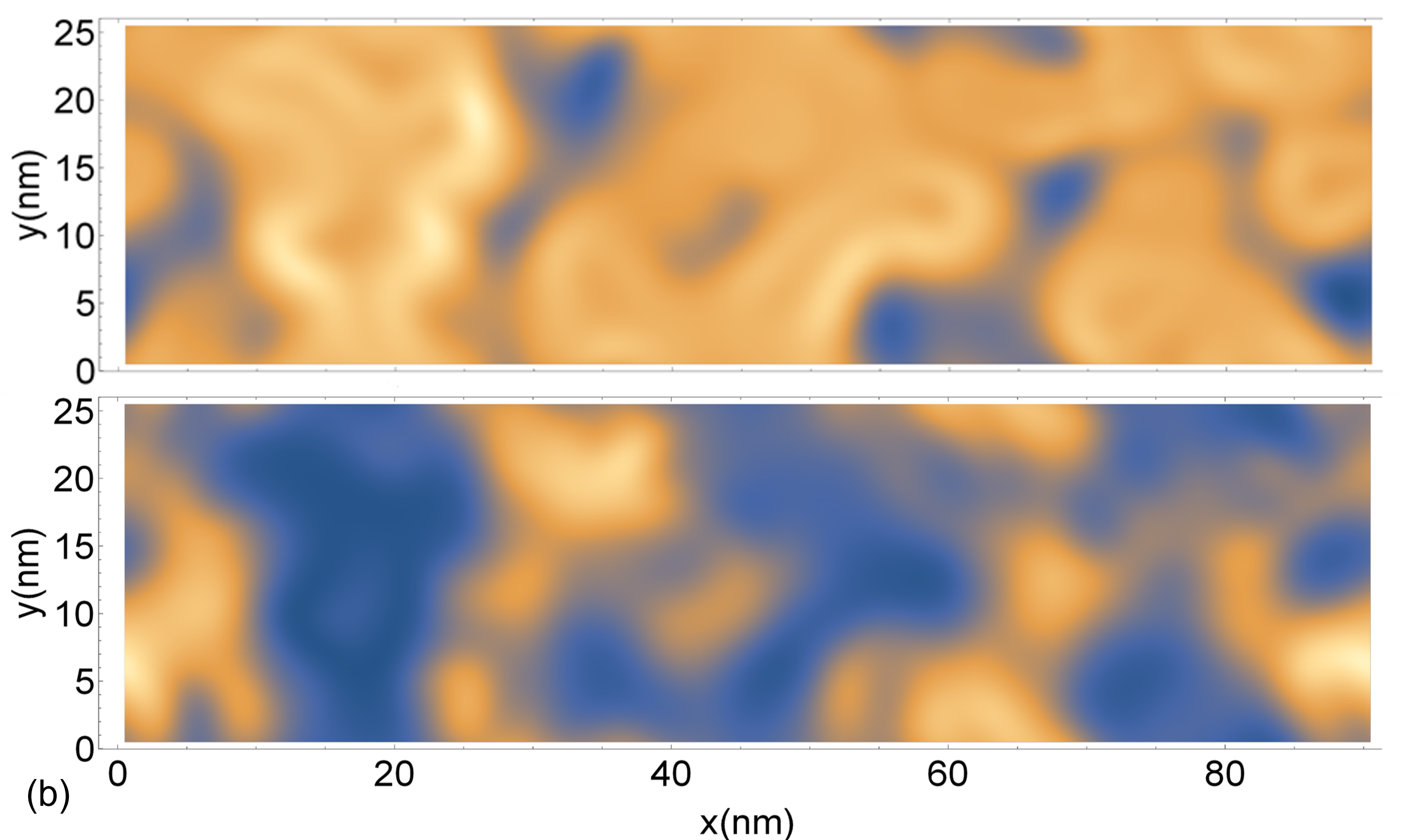}\label{fig:2D_LDOS(E=0)}\\
    \includegraphics[width=\columnwidth]{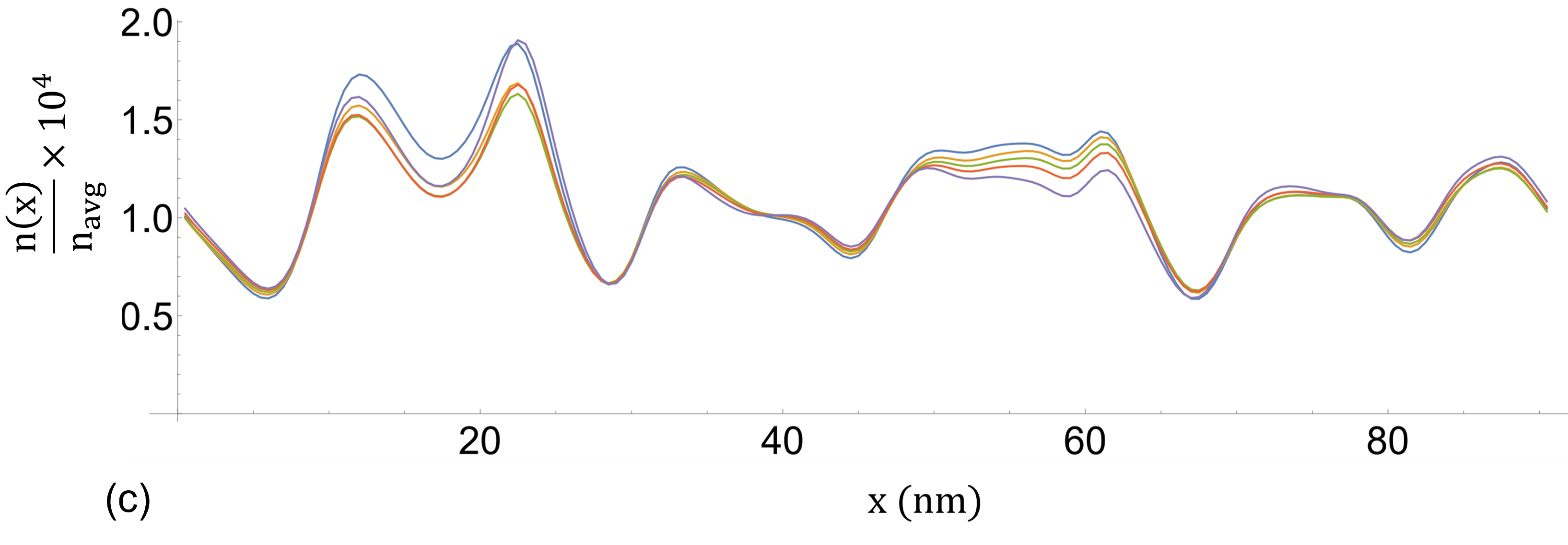}\label{fig:2D_LDOS_section}   
    \caption{ (a) Trivial islands shown as white patches whereas dark blue region is topological, (b) top layer 2D LDOS and next layer 2D LDOS shown in colour map. (c) Normalized LDOS plot at five different energy values along the green line as marked on the top layer as shown in (a)}\label{fig:2D_topodisorder}
\end{figure}
The results discussed so far have been based on a simplified model of disorder where the variation of the  disorder parameter along the $y$-direction was ignored. Finally, we present the results from the simulation of a more realistic model where the topological domain disorder is introduced in 2D at each layer. The corresponding disorder potential in momentum space, $w_{2D}(k_x,k_y) $, can be obtained by substituting $k_x $ in Eq. \ref{eq:disorderdefn} by $k_{\parallel} = \sqrt{k_x^2+k_y^2} $. In this case the trivial islands (shown as white regions on the top layer in \ref{fig:2D_topodisorder}(a)) are finite in size along all directions. It can be noted from the colour plots for the 2D LDOS ( Fig.\ref{fig:2D_topodisorder} (b)) and correlating to the trivial island picture that at the trivial regions the top layer LDOS is reduced whereas the second layer LDOS is increased. This is consistent with the result for the 1D disorder case as discussed before. Moreover, the energy collapse feature of the LDOS at different energies (shown for a section of the sample on the top layer in Fig. \ref{fig:2D_topodisorder}(c)) is also present in the 2D disordered case.
\section{Summary and discussion\label{sec:conclusion}}
In this work, we have studied the effect of alloy disorder (i.e. Te/Se composition fluctuations) on the topological surface state of FTS. In order to do this, we have introduced a variant of the BHZ model \cite{BHZmodel} for the strong topological insulator phase that is appropriate to a layered system such as FTS where the bandwidth from out-of-plane tunneling is much smaller than the in-plane bandwidths. Similar 
to the BHZ model, the topological properties such as the surface state dispersion and the near-gap bulk band structure can be characterized with a few phenomenologically determined parameters. We have fit this model to ARPES \cite{KanigelARPES,ShikSinARPES} and been able to 
reproduce the qualitative structure of the surface ARPES dispersion. The BHZ model, which is based on a small wave-vector approximation, is not appropriate to FTS because of the narrow band width in the out of plane direction. Our phenomenological model can be combined with input from DFT calculations\cite{ZhongFang} to model the effect of Te composition fluctuations as a shift of the odd parity $p_z $ band. We find that fluctuations in the position of the odd parity band can drive local fluctuations in the topological invariant, which in turn can lead to the disappearance of the surface state in parts 
of the surface. The disappearance of the surface state is marked by a reduced local density of states on the top surface in these areas, which are bounded by domain-walls of peaks in the LDOS. One complication that we discuss is that the suppressed density of states in the non-topological domains may appear similar to topological areas where the Fermi-level crosses the Dirac point. We have shown that the 
effects of topological domain disorder can be distinguished from other forms of disorder by considering the energy dependence of the pattern of the fluctuation of density of states (also called quasiparticle interference).

As mentioned in the introduction, our motivation to study the effect of alloy disorder on the topological surface state was based on its potential effect on the vortex MBSs. As an example of how local fluctuations in the topological character of a system can affect MBSs, consider a scenario where in the SC state a vortex core penetrates through a trivial region on the top surface. Since the trivial region does not support topological states on the surface, such a vortex would not be expected to support any MBS. This would in principle be a potential explanation for the absence of MBSs in a large fraction of vortex cores in experiments \cite{HanaguriZBPs,WenHanagurilike}. Furthermore, such a topological domain disordered phase might also help explain the absence of quantized conductance in a large set of vortices\cite{zhu2020nearly}. This would occur when the trivial region is small in size relative to the superconducting coherence length so that the MBS wave-function is still accessible to tunneling but at a reduced tunneling strength. Quantitative effect on MBS wave-function would depend on the actual correlation length for the alloy disorder and must be left to future work.
 
The alloy disorder that is natural in FTS can have other effects such as change in band width of the odd parity band and the topological gap. In fact, the disappearance of the topological surface state in experiments~\cite{Brookhaven} can in principle arise from a reduction or shift
in the topological gap as opposed to a change in the topological invariant discussed in the manuscript. However, since DFT calculations or experimental data are not available to estimate these effects making this difficult to estimate at present. Furthermore, correlation between the alloy disorder i.e. Te/Se positions may complicate the analysis of the disorder effects.
However, one can hope that these effects together with iron impurity induced chemical potential fluctuations~\cite{GhaemiZeemancoupltheory} as well as nematic fluctuations and strain effects can be included into the surface state model to develop a complete understanding of the evolution of vortex level spectra with magnetic field~\cite{HanaguriZBPs}.

This work was supported by NSF QII-TAQS 1936246. JDS would also like to thank NSF DMR-1555135 (CAREER) for support. We also thank  Jinying Wang, Jennifer Hoffman, Pouyan Ghaemi and Ruixing Zhang for valuable discussions.

\bigskip
\bibliography{main}

\begin{thebibliography}{28}%
\makeatletter
\providecommand \@ifxundefined [1]{%
 \@ifx{#1\undefined}
}%
\providecommand \@ifnum [1]{%
 \ifnum #1\expandafter \@firstoftwo
 \else \expandafter \@secondoftwo
 \fi
}%
\providecommand \@ifx [1]{%
 \ifx #1\expandafter \@firstoftwo
 \else \expandafter \@secondoftwo
 \fi
}%
\providecommand \natexlab [1]{#1}%
\providecommand \enquote  [1]{``#1''}%
\providecommand \bibnamefont  [1]{#1}%
\providecommand \bibfnamefont [1]{#1}%
\providecommand \citenamefont [1]{#1}%
\providecommand \href@noop [0]{\@secondoftwo}%
\providecommand \href [0]{\begingroup \@sanitize@url \@href}%
\providecommand \@href[1]{\@@startlink{#1}\@@href}%
\providecommand \@@href[1]{\endgroup#1\@@endlink}%
\providecommand \@sanitize@url [0]{\catcode `\\12\catcode `\$12\catcode
  `\&12\catcode `\#12\catcode `\^12\catcode `\_12\catcode `\%12\relax}%
\providecommand \@@startlink[1]{}%
\providecommand \@@endlink[0]{}%
\providecommand \url  [0]{\begingroup\@sanitize@url \@url }%
\providecommand \@url [1]{\endgroup\@href {#1}{\urlprefix }}%
\providecommand \urlprefix  [0]{URL }%
\providecommand \Eprint [0]{\href }%
\providecommand \doibase [0]{http://dx.doi.org/}%
\providecommand \selectlanguage [0]{\@gobble}%
\providecommand \bibinfo  [0]{\@secondoftwo}%
\providecommand \bibfield  [0]{\@secondoftwo}%
\providecommand \translation [1]{[#1]}%
\providecommand \BibitemOpen [0]{}%
\providecommand \bibitemStop [0]{}%
\providecommand \bibitemNoStop [0]{.\EOS\space}%
\providecommand \EOS [0]{\spacefactor3000\relax}%
\providecommand \BibitemShut  [1]{\csname bibitem#1\endcsname}%
\let\auto@bib@innerbib\@empty
\bibitem [{\citenamefont {Wang}\ \emph {et~al.}(2015)\citenamefont {Wang},
  \citenamefont {Zhang}, \citenamefont {Xu}, \citenamefont {Zeng},
  \citenamefont {Miao}, \citenamefont {Xu}, \citenamefont {Qian}, \citenamefont
  {Weng}, \citenamefont {Richard}, \citenamefont {Fedorov}, \citenamefont
  {Ding}, \citenamefont {Dai},\ and\ \citenamefont {Fang}}]{ZhongFang}%
  \BibitemOpen
  \bibfield  {author} {\bibinfo {author} {\bibfnamefont {Z.}~\bibnamefont
  {Wang}}, \bibinfo {author} {\bibfnamefont {P.}~\bibnamefont {Zhang}},
  \bibinfo {author} {\bibfnamefont {G.}~\bibnamefont {Xu}}, \bibinfo {author}
  {\bibfnamefont {L.~K.}\ \bibnamefont {Zeng}}, \bibinfo {author}
  {\bibfnamefont {H.}~\bibnamefont {Miao}}, \bibinfo {author} {\bibfnamefont
  {X.}~\bibnamefont {Xu}}, \bibinfo {author} {\bibfnamefont {T.}~\bibnamefont
  {Qian}}, \bibinfo {author} {\bibfnamefont {H.}~\bibnamefont {Weng}}, \bibinfo
  {author} {\bibfnamefont {P.}~\bibnamefont {Richard}}, \bibinfo {author}
  {\bibfnamefont {A.~V.}\ \bibnamefont {Fedorov}}, \bibinfo {author}
  {\bibfnamefont {H.}~\bibnamefont {Ding}}, \bibinfo {author} {\bibfnamefont
  {X.}~\bibnamefont {Dai}}, \ and\ \bibinfo {author} {\bibfnamefont
  {Z.}~\bibnamefont {Fang}},\ }\href {\doibase 10.1103/PhysRevB.92.115119}
  {\bibfield  {journal} {\bibinfo  {journal} {Physical Review B}\ }\textbf
  {\bibinfo {volume} {92}} (\bibinfo {year} {2015}),\
  10.1103/PhysRevB.92.115119}\BibitemShut {NoStop}%
\bibitem [{\citenamefont {Xu}\ \emph {et~al.}(2016)\citenamefont {Xu},
  \citenamefont {Lian}, \citenamefont {Tang}, \citenamefont {Qi},\ and\
  \citenamefont {Zhang}}]{SCZhang_TSC}%
  \BibitemOpen
  \bibfield  {author} {\bibinfo {author} {\bibfnamefont {G.}~\bibnamefont
  {Xu}}, \bibinfo {author} {\bibfnamefont {B.}~\bibnamefont {Lian}}, \bibinfo
  {author} {\bibfnamefont {P.}~\bibnamefont {Tang}}, \bibinfo {author}
  {\bibfnamefont {X.-L.}\ \bibnamefont {Qi}}, \ and\ \bibinfo {author}
  {\bibfnamefont {S.-C.}\ \bibnamefont {Zhang}},\ }\href {\doibase
  10.1103/PhysRevLett.117.047001} {\bibfield  {journal} {\bibinfo  {journal}
  {Physical Review Letters}\ }\textbf {\bibinfo {volume} {117}} (\bibinfo
  {year} {2016}),\ 10.1103/PhysRevLett.117.047001}\BibitemShut {NoStop}%
\bibitem [{\citenamefont {Lohani}\ \emph {et~al.}(2020)\citenamefont {Lohani},
  \citenamefont {Hazra}, \citenamefont {Ribak}, \citenamefont {Nitzav},
  \citenamefont {Fu}, \citenamefont {Yan}, \citenamefont {Randeria},\ and\
  \citenamefont {Kanigel}}]{KanigelARPES}%
  \BibitemOpen
  \bibfield  {author} {\bibinfo {author} {\bibfnamefont {H.}~\bibnamefont
  {Lohani}}, \bibinfo {author} {\bibfnamefont {T.}~\bibnamefont {Hazra}},
  \bibinfo {author} {\bibfnamefont {A.}~\bibnamefont {Ribak}}, \bibinfo
  {author} {\bibfnamefont {Y.}~\bibnamefont {Nitzav}}, \bibinfo {author}
  {\bibfnamefont {H.}~\bibnamefont {Fu}}, \bibinfo {author} {\bibfnamefont
  {B.}~\bibnamefont {Yan}}, \bibinfo {author} {\bibfnamefont {M.}~\bibnamefont
  {Randeria}}, \ and\ \bibinfo {author} {\bibfnamefont {A.}~\bibnamefont
  {Kanigel}},\ }\href {\doibase 10.1103/PhysRevB.101.245146} {\bibfield
  {journal} {\bibinfo  {journal} {Phys. Rev. B}\ }\textbf {\bibinfo {volume}
  {101}},\ \bibinfo {pages} {245146} (\bibinfo {year} {2020})}\BibitemShut
  {NoStop}%
\bibitem [{\citenamefont {Zhang}\ \emph {et~al.}(2018)\citenamefont {Zhang},
  \citenamefont {Yaji}, \citenamefont {Hashimoto}, \citenamefont {Ota},
  \citenamefont {Kondo}, \citenamefont {Okazaki}, \citenamefont {Wang},
  \citenamefont {Wen}, \citenamefont {Gu}, \citenamefont {Ding},\ and\
  \citenamefont {Shin}}]{ShikSinARPES}%
  \BibitemOpen
  \bibfield  {author} {\bibinfo {author} {\bibfnamefont {P.}~\bibnamefont
  {Zhang}}, \bibinfo {author} {\bibfnamefont {K.}~\bibnamefont {Yaji}},
  \bibinfo {author} {\bibfnamefont {T.}~\bibnamefont {Hashimoto}}, \bibinfo
  {author} {\bibfnamefont {Y.}~\bibnamefont {Ota}}, \bibinfo {author}
  {\bibfnamefont {T.}~\bibnamefont {Kondo}}, \bibinfo {author} {\bibfnamefont
  {K.}~\bibnamefont {Okazaki}}, \bibinfo {author} {\bibfnamefont
  {Z.}~\bibnamefont {Wang}}, \bibinfo {author} {\bibfnamefont {J.}~\bibnamefont
  {Wen}}, \bibinfo {author} {\bibfnamefont {G.~D.}\ \bibnamefont {Gu}},
  \bibinfo {author} {\bibfnamefont {H.}~\bibnamefont {Ding}}, \ and\ \bibinfo
  {author} {\bibfnamefont {S.}~\bibnamefont {Shin}},\ }\href {\doibase
  10.1126/science.aan4596} {\bibfield  {journal} {\bibinfo  {journal}
  {Science}\ }\textbf {\bibinfo {volume} {360}},\ \bibinfo {pages} {182}
  (\bibinfo {year} {2018})}\BibitemShut {NoStop}%
\bibitem [{\citenamefont {Zhang}\ \emph {et~al.}(2019)\citenamefont {Zhang},
  \citenamefont {Wang}, \citenamefont {Wu}, \citenamefont {Yaji}, \citenamefont
  {Ishida}, \citenamefont {Kohama}, \citenamefont {Dai}, \citenamefont {Sun},
  \citenamefont {Bareille}, \citenamefont {Kuroda}, \citenamefont {Kondo},
  \citenamefont {Okazaki}, \citenamefont {Kindo}, \citenamefont {Wang},
  \citenamefont {Jin}, \citenamefont {Hu}, \citenamefont {Thomale},
  \citenamefont {Sumida}, \citenamefont {Wu}, \citenamefont {Miyamoto},
  \citenamefont {Okuda}, \citenamefont {Ding}, \citenamefont {Gu},
  \citenamefont {Tamegai}, \citenamefont {Kawakami}, \citenamefont {Sato},\
  and\ \citenamefont {Shin}}]{Zhang2019}%
  \BibitemOpen
  \bibfield  {author} {\bibinfo {author} {\bibfnamefont {P.}~\bibnamefont
  {Zhang}}, \bibinfo {author} {\bibfnamefont {Z.}~\bibnamefont {Wang}},
  \bibinfo {author} {\bibfnamefont {X.}~\bibnamefont {Wu}}, \bibinfo {author}
  {\bibfnamefont {K.}~\bibnamefont {Yaji}}, \bibinfo {author} {\bibfnamefont
  {Y.}~\bibnamefont {Ishida}}, \bibinfo {author} {\bibfnamefont
  {Y.}~\bibnamefont {Kohama}}, \bibinfo {author} {\bibfnamefont
  {G.}~\bibnamefont {Dai}}, \bibinfo {author} {\bibfnamefont {Y.}~\bibnamefont
  {Sun}}, \bibinfo {author} {\bibfnamefont {C.}~\bibnamefont {Bareille}},
  \bibinfo {author} {\bibfnamefont {K.}~\bibnamefont {Kuroda}}, \bibinfo
  {author} {\bibfnamefont {T.}~\bibnamefont {Kondo}}, \bibinfo {author}
  {\bibfnamefont {K.}~\bibnamefont {Okazaki}}, \bibinfo {author} {\bibfnamefont
  {K.}~\bibnamefont {Kindo}}, \bibinfo {author} {\bibfnamefont
  {X.}~\bibnamefont {Wang}}, \bibinfo {author} {\bibfnamefont {C.}~\bibnamefont
  {Jin}}, \bibinfo {author} {\bibfnamefont {J.}~\bibnamefont {Hu}}, \bibinfo
  {author} {\bibfnamefont {R.}~\bibnamefont {Thomale}}, \bibinfo {author}
  {\bibfnamefont {K.}~\bibnamefont {Sumida}}, \bibinfo {author} {\bibfnamefont
  {S.}~\bibnamefont {Wu}}, \bibinfo {author} {\bibfnamefont {K.}~\bibnamefont
  {Miyamoto}}, \bibinfo {author} {\bibfnamefont {T.}~\bibnamefont {Okuda}},
  \bibinfo {author} {\bibfnamefont {H.}~\bibnamefont {Ding}}, \bibinfo {author}
  {\bibfnamefont {G.~D.}\ \bibnamefont {Gu}}, \bibinfo {author} {\bibfnamefont
  {T.}~\bibnamefont {Tamegai}}, \bibinfo {author} {\bibfnamefont
  {T.}~\bibnamefont {Kawakami}}, \bibinfo {author} {\bibfnamefont
  {M.}~\bibnamefont {Sato}}, \ and\ \bibinfo {author} {\bibfnamefont
  {S.}~\bibnamefont {Shin}},\ }\href {\doibase 10.1038/s41567-018-0280-z}
  {\bibfield  {journal} {\bibinfo  {journal} {Nature Physics}\ }\textbf
  {\bibinfo {volume} {15}} (\bibinfo {year} {2019}),\
  10.1038/s41567-018-0280-z}\BibitemShut {NoStop}%
\bibitem [{\citenamefont {Fu}\ and\ \citenamefont {Kane}(2008)}]{KaneMZM}%
  \BibitemOpen
  \bibfield  {author} {\bibinfo {author} {\bibfnamefont {L.}~\bibnamefont
  {Fu}}\ and\ \bibinfo {author} {\bibfnamefont {C.~L.}\ \bibnamefont {Kane}},\
  }\href {\doibase 10.1103/PhysRevLett.100.096407} {\bibfield  {journal}
  {\bibinfo  {journal} {Phys. Rev. Lett.}\ }\textbf {\bibinfo {volume} {100}},\
  \bibinfo {pages} {096407} (\bibinfo {year} {2008})}\BibitemShut {NoStop}%
\bibitem [{\citenamefont {Hosur}\ \emph {et~al.}(2011)\citenamefont {Hosur},
  \citenamefont {Ghaemi}, \citenamefont {Mong},\ and\ \citenamefont
  {Vishwanath}}]{hosur2011majorana}%
  \BibitemOpen
  \bibfield  {author} {\bibinfo {author} {\bibfnamefont {P.}~\bibnamefont
  {Hosur}}, \bibinfo {author} {\bibfnamefont {P.}~\bibnamefont {Ghaemi}},
  \bibinfo {author} {\bibfnamefont {R.~S.~K.}\ \bibnamefont {Mong}}, \ and\
  \bibinfo {author} {\bibfnamefont {A.}~\bibnamefont {Vishwanath}},\ }\href
  {\doibase 10.1103/PhysRevLett.107.097001} {\bibfield  {journal} {\bibinfo
  {journal} {Phys. Rev. Lett.}\ }\textbf {\bibinfo {volume} {107}},\ \bibinfo
  {pages} {097001} (\bibinfo {year} {2011})}\BibitemShut {NoStop}%
\bibitem [{\citenamefont {Xu}\ \emph {et~al.}(2015)\citenamefont {Xu},
  \citenamefont {Wang}, \citenamefont {Liu}, \citenamefont {Ge}, \citenamefont
  {Yang}, \citenamefont {Liu}, \citenamefont {Xu}, \citenamefont {Guan},
  \citenamefont {Gao}, \citenamefont {Qian}, \citenamefont {Liu}, \citenamefont
  {Wang}, \citenamefont {Zhang}, \citenamefont {Xue},\ and\ \citenamefont
  {Jia}}]{MZMinBiTe}%
  \BibitemOpen
  \bibfield  {author} {\bibinfo {author} {\bibfnamefont {J.-P.}\ \bibnamefont
  {Xu}}, \bibinfo {author} {\bibfnamefont {M.-X.}\ \bibnamefont {Wang}},
  \bibinfo {author} {\bibfnamefont {Z.~L.}\ \bibnamefont {Liu}}, \bibinfo
  {author} {\bibfnamefont {J.-F.}\ \bibnamefont {Ge}}, \bibinfo {author}
  {\bibfnamefont {X.}~\bibnamefont {Yang}}, \bibinfo {author} {\bibfnamefont
  {C.}~\bibnamefont {Liu}}, \bibinfo {author} {\bibfnamefont {Z.~A.}\
  \bibnamefont {Xu}}, \bibinfo {author} {\bibfnamefont {D.}~\bibnamefont
  {Guan}}, \bibinfo {author} {\bibfnamefont {C.~L.}\ \bibnamefont {Gao}},
  \bibinfo {author} {\bibfnamefont {D.}~\bibnamefont {Qian}}, \bibinfo {author}
  {\bibfnamefont {Y.}~\bibnamefont {Liu}}, \bibinfo {author} {\bibfnamefont
  {Q.-H.}\ \bibnamefont {Wang}}, \bibinfo {author} {\bibfnamefont {F.-C.}\
  \bibnamefont {Zhang}}, \bibinfo {author} {\bibfnamefont {Q.-K.}\ \bibnamefont
  {Xue}}, \ and\ \bibinfo {author} {\bibfnamefont {J.-F.}\ \bibnamefont
  {Jia}},\ }\href {\doibase 10.1103/PhysRevLett.114.017001} {\bibfield
  {journal} {\bibinfo  {journal} {Phys. Rev. Lett.}\ }\textbf {\bibinfo
  {volume} {114}},\ \bibinfo {pages} {017001} (\bibinfo {year}
  {2015})}\BibitemShut {NoStop}%
\bibitem [{\citenamefont {Sun}\ \emph {et~al.}(2016)\citenamefont {Sun},
  \citenamefont {Zhang}, \citenamefont {Hu}, \citenamefont {Li}, \citenamefont
  {Wang}, \citenamefont {Ma}, \citenamefont {Xu}, \citenamefont {Gao},
  \citenamefont {Guan}, \citenamefont {Li}, \citenamefont {Liu}, \citenamefont
  {Qian}, \citenamefont {Zhou}, \citenamefont {Fu}, \citenamefont {Li},
  \citenamefont {Zhang},\ and\ \citenamefont {Jia}}]{MZMSSARinBiTe}%
  \BibitemOpen
  \bibfield  {author} {\bibinfo {author} {\bibfnamefont {H.-H.}\ \bibnamefont
  {Sun}}, \bibinfo {author} {\bibfnamefont {K.-W.}\ \bibnamefont {Zhang}},
  \bibinfo {author} {\bibfnamefont {L.-H.}\ \bibnamefont {Hu}}, \bibinfo
  {author} {\bibfnamefont {C.}~\bibnamefont {Li}}, \bibinfo {author}
  {\bibfnamefont {G.-Y.}\ \bibnamefont {Wang}}, \bibinfo {author}
  {\bibfnamefont {H.-Y.}\ \bibnamefont {Ma}}, \bibinfo {author} {\bibfnamefont
  {Z.-A.}\ \bibnamefont {Xu}}, \bibinfo {author} {\bibfnamefont {C.-L.}\
  \bibnamefont {Gao}}, \bibinfo {author} {\bibfnamefont {D.-D.}\ \bibnamefont
  {Guan}}, \bibinfo {author} {\bibfnamefont {Y.-Y.}\ \bibnamefont {Li}},
  \bibinfo {author} {\bibfnamefont {C.}~\bibnamefont {Liu}}, \bibinfo {author}
  {\bibfnamefont {D.}~\bibnamefont {Qian}}, \bibinfo {author} {\bibfnamefont
  {Y.}~\bibnamefont {Zhou}}, \bibinfo {author} {\bibfnamefont {L.}~\bibnamefont
  {Fu}}, \bibinfo {author} {\bibfnamefont {S.-C.}\ \bibnamefont {Li}}, \bibinfo
  {author} {\bibfnamefont {F.-C.}\ \bibnamefont {Zhang}}, \ and\ \bibinfo
  {author} {\bibfnamefont {J.-F.}\ \bibnamefont {Jia}},\ }\href {\doibase
  10.1103/PhysRevLett.116.257003} {\bibfield  {journal} {\bibinfo  {journal}
  {Phys. Rev. Lett.}\ }\textbf {\bibinfo {volume} {116}},\ \bibinfo {pages}
  {257003} (\bibinfo {year} {2016})}\BibitemShut {NoStop}%
\bibitem [{\citenamefont {Liu}\ \emph {et~al.}(2018)\citenamefont {Liu},
  \citenamefont {Chen}, \citenamefont {Zhang}, \citenamefont {Peng},
  \citenamefont {Yan}, \citenamefont {Wen}, \citenamefont {Lou}, \citenamefont
  {Huang}, \citenamefont {Tian}, \citenamefont {Dong}, \citenamefont {Wang},
  \citenamefont {Bao}, \citenamefont {Wang}, \citenamefont {Yin}, \citenamefont
  {Zhao},\ and\ \citenamefont {Feng}}]{MZMinLiFeSe}%
  \BibitemOpen
  \bibfield  {author} {\bibinfo {author} {\bibfnamefont {Q.}~\bibnamefont
  {Liu}}, \bibinfo {author} {\bibfnamefont {C.}~\bibnamefont {Chen}}, \bibinfo
  {author} {\bibfnamefont {T.}~\bibnamefont {Zhang}}, \bibinfo {author}
  {\bibfnamefont {R.}~\bibnamefont {Peng}}, \bibinfo {author} {\bibfnamefont
  {Y.-J.}\ \bibnamefont {Yan}}, \bibinfo {author} {\bibfnamefont {C.-H.-P.}\
  \bibnamefont {Wen}}, \bibinfo {author} {\bibfnamefont {X.}~\bibnamefont
  {Lou}}, \bibinfo {author} {\bibfnamefont {Y.-L.}\ \bibnamefont {Huang}},
  \bibinfo {author} {\bibfnamefont {J.-P.}\ \bibnamefont {Tian}}, \bibinfo
  {author} {\bibfnamefont {X.-L.}\ \bibnamefont {Dong}}, \bibinfo {author}
  {\bibfnamefont {G.-W.}\ \bibnamefont {Wang}}, \bibinfo {author}
  {\bibfnamefont {W.-C.}\ \bibnamefont {Bao}}, \bibinfo {author} {\bibfnamefont
  {Q.-H.}\ \bibnamefont {Wang}}, \bibinfo {author} {\bibfnamefont {Z.-P.}\
  \bibnamefont {Yin}}, \bibinfo {author} {\bibfnamefont {Z.-X.}\ \bibnamefont
  {Zhao}}, \ and\ \bibinfo {author} {\bibfnamefont {D.-L.}\ \bibnamefont
  {Feng}},\ }\href {\doibase 10.1103/PhysRevX.8.041056} {\bibfield  {journal}
  {\bibinfo  {journal} {Phys. Rev. X}\ }\textbf {\bibinfo {volume} {8}},\
  \bibinfo {pages} {041056} (\bibinfo {year} {2018})}\BibitemShut {NoStop}%
\bibitem [{\citenamefont {Kitaev}(2003)}]{KitaevQC}%
  \BibitemOpen
  \bibfield  {author} {\bibinfo {author} {\bibfnamefont {A.}~\bibnamefont
  {Kitaev}},\ }\href {\doibase 10.1016/S0003-4916(02)00018-0} {\bibfield
  {journal} {\bibinfo  {journal} {Annals of Physics}\ }\textbf {\bibinfo
  {volume} {303}} (\bibinfo {year} {2003}),\
  10.1016/S0003-4916(02)00018-0}\BibitemShut {NoStop}%
\bibitem [{\citenamefont {Nayak}\ \emph {et~al.}(2008)\citenamefont {Nayak},
  \citenamefont {Simon}, \citenamefont {Stern}, \citenamefont {Freedman},\ and\
  \citenamefont {Das~Sarma}}]{NayakTQCrev}%
  \BibitemOpen
  \bibfield  {author} {\bibinfo {author} {\bibfnamefont {C.}~\bibnamefont
  {Nayak}}, \bibinfo {author} {\bibfnamefont {S.~H.}\ \bibnamefont {Simon}},
  \bibinfo {author} {\bibfnamefont {A.}~\bibnamefont {Stern}}, \bibinfo
  {author} {\bibfnamefont {M.}~\bibnamefont {Freedman}}, \ and\ \bibinfo
  {author} {\bibfnamefont {S.}~\bibnamefont {Das~Sarma}},\ }\href {\doibase
  10.1103/RevModPhys.80.1083} {\bibfield  {journal} {\bibinfo  {journal} {Rev.
  Mod. Phys.}\ }\textbf {\bibinfo {volume} {80}},\ \bibinfo {pages} {1083}
  (\bibinfo {year} {2008})}\BibitemShut {NoStop}%
\bibitem [{\citenamefont {Sato}\ and\ \citenamefont {Ando}(2017)}]{Sato2017}%
  \BibitemOpen
  \bibfield  {author} {\bibinfo {author} {\bibfnamefont {M.}~\bibnamefont
  {Sato}}\ and\ \bibinfo {author} {\bibfnamefont {Y.}~\bibnamefont {Ando}},\
  }\href {\doibase 10.1088/1361-6633/aa6ac7} {\bibfield  {journal} {\bibinfo
  {journal} {Reports on Progress in Physics}\ }\textbf {\bibinfo {volume}
  {80}},\ \bibinfo {pages} {076501} (\bibinfo {year} {2017})}\BibitemShut
  {NoStop}%
\bibitem [{\citenamefont {Machida}\ \emph {et~al.}(2019)\citenamefont
  {Machida}, \citenamefont {Sun}, \citenamefont {Pyon}, \citenamefont {Takeda},
  \citenamefont {Kohsaka}, \citenamefont {Hanaguri}, \citenamefont {Sasagawa},\
  and\ \citenamefont {Tamegai}}]{HanaguriZBPs}%
  \BibitemOpen
  \bibfield  {author} {\bibinfo {author} {\bibfnamefont {T.}~\bibnamefont
  {Machida}}, \bibinfo {author} {\bibfnamefont {Y.}~\bibnamefont {Sun}},
  \bibinfo {author} {\bibfnamefont {S.}~\bibnamefont {Pyon}}, \bibinfo {author}
  {\bibfnamefont {S.}~\bibnamefont {Takeda}}, \bibinfo {author} {\bibfnamefont
  {Y.}~\bibnamefont {Kohsaka}}, \bibinfo {author} {\bibfnamefont
  {T.}~\bibnamefont {Hanaguri}}, \bibinfo {author} {\bibfnamefont
  {T.}~\bibnamefont {Sasagawa}}, \ and\ \bibinfo {author} {\bibfnamefont
  {T.}~\bibnamefont {Tamegai}},\ }\href {\doibase 10.1038/s41563-019-0397-1}
  {\bibfield  {journal} {\bibinfo  {journal} {Nature Materials}\ }\textbf
  {\bibinfo {volume} {18}} (\bibinfo {year} {2019}),\
  10.1038/s41563-019-0397-1}\BibitemShut {NoStop}%
\bibitem [{\citenamefont {Wang}\ \emph {et~al.}(2018)\citenamefont {Wang},
  \citenamefont {Kong}, \citenamefont {Fan}, \citenamefont {Chen},
  \citenamefont {Zhu}, \citenamefont {Liu}, \citenamefont {Cao}, \citenamefont
  {Sun}, \citenamefont {Du}, \citenamefont {Schneeloch}, \citenamefont {Zhong},
  \citenamefont {Gu}, \citenamefont {Fu}, \citenamefont {Ding},\ and\
  \citenamefont {Gao}}]{Dongfeietal}%
  \BibitemOpen
  \bibfield  {author} {\bibinfo {author} {\bibfnamefont {D.}~\bibnamefont
  {Wang}}, \bibinfo {author} {\bibfnamefont {L.}~\bibnamefont {Kong}}, \bibinfo
  {author} {\bibfnamefont {P.}~\bibnamefont {Fan}}, \bibinfo {author}
  {\bibfnamefont {H.}~\bibnamefont {Chen}}, \bibinfo {author} {\bibfnamefont
  {S.}~\bibnamefont {Zhu}}, \bibinfo {author} {\bibfnamefont {W.}~\bibnamefont
  {Liu}}, \bibinfo {author} {\bibfnamefont {L.}~\bibnamefont {Cao}}, \bibinfo
  {author} {\bibfnamefont {Y.}~\bibnamefont {Sun}}, \bibinfo {author}
  {\bibfnamefont {S.}~\bibnamefont {Du}}, \bibinfo {author} {\bibfnamefont
  {J.}~\bibnamefont {Schneeloch}}, \bibinfo {author} {\bibfnamefont
  {R.}~\bibnamefont {Zhong}}, \bibinfo {author} {\bibfnamefont
  {G.}~\bibnamefont {Gu}}, \bibinfo {author} {\bibfnamefont {L.}~\bibnamefont
  {Fu}}, \bibinfo {author} {\bibfnamefont {H.}~\bibnamefont {Ding}}, \ and\
  \bibinfo {author} {\bibfnamefont {H.-J.}\ \bibnamefont {Gao}},\ }\href
  {\doibase 10.1126/science.aao1797} {\bibfield  {journal} {\bibinfo  {journal}
  {Science}\ }\textbf {\bibinfo {volume} {362}} (\bibinfo {year} {2018}),\
  10.1126/science.aao1797}\BibitemShut {NoStop}%
\bibitem [{\citenamefont {Chen}\ \emph {et~al.}(2019)\citenamefont {Chen},
  \citenamefont {Chen}, \citenamefont {Duan}, \citenamefont {Zhu},
  \citenamefont {Yang},\ and\ \citenamefont {Wen}}]{WenHanagurilike}%
  \BibitemOpen
  \bibfield  {author} {\bibinfo {author} {\bibfnamefont {X.}~\bibnamefont
  {Chen}}, \bibinfo {author} {\bibfnamefont {M.}~\bibnamefont {Chen}}, \bibinfo
  {author} {\bibfnamefont {W.}~\bibnamefont {Duan}}, \bibinfo {author}
  {\bibfnamefont {X.}~\bibnamefont {Zhu}}, \bibinfo {author} {\bibfnamefont
  {H.}~\bibnamefont {Yang}}, \ and\ \bibinfo {author} {\bibfnamefont {H.-H.}\
  \bibnamefont {Wen}},\ }\href {http://arxiv.org/abs/1909.01686} {\bibfield
  {journal} {\bibinfo  {journal} {http://arxiv.org/abs/1909.01686}\ } (\bibinfo
  {year} {2019})}\BibitemShut {NoStop}%
\bibitem [{\citenamefont {Kong}\ \emph {et~al.}(2019)\citenamefont {Kong},
  \citenamefont {Zhu}, \citenamefont {Papaj}, \citenamefont {Chen},
  \citenamefont {Cao}, \citenamefont {Isobe}, \citenamefont {Xing},
  \citenamefont {Liu}, \citenamefont {Wang}, \citenamefont {Fan}, \citenamefont
  {Sun}, \citenamefont {Du}, \citenamefont {Schneeloch}, \citenamefont {Zhong},
  \citenamefont {Gu}, \citenamefont {Fu}, \citenamefont {Gao},\ and\
  \citenamefont {Ding}}]{CdGM_FTS_2019}%
  \BibitemOpen
  \bibfield  {author} {\bibinfo {author} {\bibfnamefont {L.}~\bibnamefont
  {Kong}}, \bibinfo {author} {\bibfnamefont {S.}~\bibnamefont {Zhu}}, \bibinfo
  {author} {\bibfnamefont {M.}~\bibnamefont {Papaj}}, \bibinfo {author}
  {\bibfnamefont {H.}~\bibnamefont {Chen}}, \bibinfo {author} {\bibfnamefont
  {L.}~\bibnamefont {Cao}}, \bibinfo {author} {\bibfnamefont {H.}~\bibnamefont
  {Isobe}}, \bibinfo {author} {\bibfnamefont {Y.}~\bibnamefont {Xing}},
  \bibinfo {author} {\bibfnamefont {W.}~\bibnamefont {Liu}}, \bibinfo {author}
  {\bibfnamefont {D.}~\bibnamefont {Wang}}, \bibinfo {author} {\bibfnamefont
  {P.}~\bibnamefont {Fan}}, \bibinfo {author} {\bibfnamefont {Y.}~\bibnamefont
  {Sun}}, \bibinfo {author} {\bibfnamefont {S.}~\bibnamefont {Du}}, \bibinfo
  {author} {\bibfnamefont {J.}~\bibnamefont {Schneeloch}}, \bibinfo {author}
  {\bibfnamefont {R.}~\bibnamefont {Zhong}}, \bibinfo {author} {\bibfnamefont
  {G.}~\bibnamefont {Gu}}, \bibinfo {author} {\bibfnamefont {L.}~\bibnamefont
  {Fu}}, \bibinfo {author} {\bibfnamefont {H.-J.}\ \bibnamefont {Gao}}, \ and\
  \bibinfo {author} {\bibfnamefont {H.}~\bibnamefont {Ding}},\ }\href {\doibase
  10.1038/s41567-019-0630-5} {\bibfield  {journal} {\bibinfo  {journal} {Nature
  Physics}\ }\textbf {\bibinfo {volume} {15}} (\bibinfo {year} {2019}),\
  10.1038/s41567-019-0630-5}\BibitemShut {NoStop}%
\bibitem [{\citenamefont {Caroli}\ \emph {et~al.}(1964)\citenamefont {Caroli},
  \citenamefont {Gennes},\ and\ \citenamefont {Matricon}}]{CdGM}%
  \BibitemOpen
  \bibfield  {author} {\bibinfo {author} {\bibfnamefont {C.}~\bibnamefont
  {Caroli}}, \bibinfo {author} {\bibfnamefont {P.~D.}\ \bibnamefont {Gennes}},
  \ and\ \bibinfo {author} {\bibfnamefont {J.}~\bibnamefont {Matricon}},\
  }\href {\doibase 10.1016/0031-9163(64)90375-0} {\bibfield  {journal}
  {\bibinfo  {journal} {Physics Letters}\ }\textbf {\bibinfo {volume} {9}}
  (\bibinfo {year} {1964}),\ 10.1016/0031-9163(64)90375-0}\BibitemShut
  {NoStop}%
\bibitem [{\citenamefont {Chen}\ \emph {et~al.}(2018)\citenamefont {Chen},
  \citenamefont {Chen}, \citenamefont {Yang}, \citenamefont {Du}, \citenamefont
  {Zhu}, \citenamefont {Wang},\ and\ \citenamefont {Wen}}]{CdGMstates}%
  \BibitemOpen
  \bibfield  {author} {\bibinfo {author} {\bibfnamefont {M.}~\bibnamefont
  {Chen}}, \bibinfo {author} {\bibfnamefont {X.}~\bibnamefont {Chen}}, \bibinfo
  {author} {\bibfnamefont {H.}~\bibnamefont {Yang}}, \bibinfo {author}
  {\bibfnamefont {Z.}~\bibnamefont {Du}}, \bibinfo {author} {\bibfnamefont
  {X.}~\bibnamefont {Zhu}}, \bibinfo {author} {\bibfnamefont {E.}~\bibnamefont
  {Wang}}, \ and\ \bibinfo {author} {\bibfnamefont {H.-H.}\ \bibnamefont
  {Wen}},\ }\href {\doibase 10.1038/s41467-018-03404-8} {\bibfield  {journal}
  {\bibinfo  {journal} {Nature Communications}\ }\textbf {\bibinfo {volume}
  {9}} (\bibinfo {year} {2018}),\ 10.1038/s41467-018-03404-8}\BibitemShut
  {NoStop}%
\bibitem [{\citenamefont {Zhu}\ \emph {et~al.}(2020)\citenamefont {Zhu},
  \citenamefont {Kong}, \citenamefont {Cao}, \citenamefont {Chen},
  \citenamefont {Papaj}, \citenamefont {Du}, \citenamefont {Xing},
  \citenamefont {Liu}, \citenamefont {Wang}, \citenamefont {Shen} \emph
  {et~al.}}]{zhu2020nearly}%
  \BibitemOpen
  \bibfield  {author} {\bibinfo {author} {\bibfnamefont {S.}~\bibnamefont
  {Zhu}}, \bibinfo {author} {\bibfnamefont {L.}~\bibnamefont {Kong}}, \bibinfo
  {author} {\bibfnamefont {L.}~\bibnamefont {Cao}}, \bibinfo {author}
  {\bibfnamefont {H.}~\bibnamefont {Chen}}, \bibinfo {author} {\bibfnamefont
  {M.}~\bibnamefont {Papaj}}, \bibinfo {author} {\bibfnamefont
  {S.}~\bibnamefont {Du}}, \bibinfo {author} {\bibfnamefont {Y.}~\bibnamefont
  {Xing}}, \bibinfo {author} {\bibfnamefont {W.}~\bibnamefont {Liu}}, \bibinfo
  {author} {\bibfnamefont {D.}~\bibnamefont {Wang}}, \bibinfo {author}
  {\bibfnamefont {C.}~\bibnamefont {Shen}},  \emph {et~al.},\ }\href
  {https://science.sciencemag.org/content/367/6474/189.abstract} {\bibfield
  {journal} {\bibinfo  {journal} {Science}\ }\textbf {\bibinfo {volume}
  {367}},\ \bibinfo {pages} {189} (\bibinfo {year} {2020})}\BibitemShut
  {NoStop}%
\bibitem [{\citenamefont {Chiu}\ \emph {et~al.}(2020)\citenamefont {Chiu},
  \citenamefont {Machida}, \citenamefont {Huang}, \citenamefont {Hanaguri},\
  and\ \citenamefont {Zhang}}]{Majoranacouplingtheory}%
  \BibitemOpen
  \bibfield  {author} {\bibinfo {author} {\bibfnamefont {C.-K.}\ \bibnamefont
  {Chiu}}, \bibinfo {author} {\bibfnamefont {T.}~\bibnamefont {Machida}},
  \bibinfo {author} {\bibfnamefont {Y.}~\bibnamefont {Huang}}, \bibinfo
  {author} {\bibfnamefont {T.}~\bibnamefont {Hanaguri}}, \ and\ \bibinfo
  {author} {\bibfnamefont {F.-C.}\ \bibnamefont {Zhang}},\ }\href {\doibase
  10.1126/sciadv.aay0443} {\bibfield  {journal} {\bibinfo  {journal} {Science
  Advances}\ }\textbf {\bibinfo {volume} {6}} (\bibinfo {year} {2020}),\
  10.1126/sciadv.aay0443}\BibitemShut {NoStop}%
\bibitem [{\citenamefont {Ghazaryan}\ \emph {et~al.}(2020)\citenamefont
  {Ghazaryan}, \citenamefont {Lopes}, \citenamefont {Hosur}, \citenamefont
  {Gilbert},\ and\ \citenamefont {Ghaemi}}]{GhaemiZeemancoupltheory}%
  \BibitemOpen
  \bibfield  {author} {\bibinfo {author} {\bibfnamefont {A.}~\bibnamefont
  {Ghazaryan}}, \bibinfo {author} {\bibfnamefont {P.~L.~S.}\ \bibnamefont
  {Lopes}}, \bibinfo {author} {\bibfnamefont {P.}~\bibnamefont {Hosur}},
  \bibinfo {author} {\bibfnamefont {M.~J.}\ \bibnamefont {Gilbert}}, \ and\
  \bibinfo {author} {\bibfnamefont {P.}~\bibnamefont {Ghaemi}},\ }\href
  {\doibase 10.1103/PhysRevB.101.020504} {\bibfield  {journal} {\bibinfo
  {journal} {Physical Review B}\ }\textbf {\bibinfo {volume} {101}} (\bibinfo
  {year} {2020}),\ 10.1103/PhysRevB.101.020504}\BibitemShut {NoStop}%
\bibitem [{\citenamefont {Li}\ \emph {et~al.}(2021)\citenamefont {Li},
  \citenamefont {Zaki}, \citenamefont {Garlea}, \citenamefont {Savici},
  \citenamefont {Fobes}, \citenamefont {Xu}, \citenamefont {Camino},
  \citenamefont {Petrovic}, \citenamefont {Gu}, \citenamefont {Johnson},
  \citenamefont {Tranquada},\ and\ \citenamefont {Zaliznyak}}]{Brookhaven}%
  \BibitemOpen
  \bibfield  {author} {\bibinfo {author} {\bibfnamefont {Y.}~\bibnamefont
  {Li}}, \bibinfo {author} {\bibfnamefont {N.}~\bibnamefont {Zaki}}, \bibinfo
  {author} {\bibfnamefont {V.~O.}\ \bibnamefont {Garlea}}, \bibinfo {author}
  {\bibfnamefont {A.~T.}\ \bibnamefont {Savici}}, \bibinfo {author}
  {\bibfnamefont {D.}~\bibnamefont {Fobes}}, \bibinfo {author} {\bibfnamefont
  {Z.}~\bibnamefont {Xu}}, \bibinfo {author} {\bibfnamefont {F.}~\bibnamefont
  {Camino}}, \bibinfo {author} {\bibfnamefont {C.}~\bibnamefont {Petrovic}},
  \bibinfo {author} {\bibfnamefont {G.}~\bibnamefont {Gu}}, \bibinfo {author}
  {\bibfnamefont {P.~D.}\ \bibnamefont {Johnson}}, \bibinfo {author}
  {\bibfnamefont {J.~M.}\ \bibnamefont {Tranquada}}, \ and\ \bibinfo {author}
  {\bibfnamefont {I.~A.}\ \bibnamefont {Zaliznyak}},\ }\href {\doibase
  10.1038/s41563-021-00984-7} {\bibfield  {journal} {\bibinfo  {journal}
  {Nature Materials}\ } (\bibinfo {year} {2021}),\
  10.1038/s41563-021-00984-7}\BibitemShut {NoStop}%
\bibitem [{\citenamefont {Hastings}\ and\ \citenamefont
  {Loring}(2011)}]{hastings2011topological}%
  \BibitemOpen
  \bibfield  {author} {\bibinfo {author} {\bibfnamefont {M.~B.}\ \bibnamefont
  {Hastings}}\ and\ \bibinfo {author} {\bibfnamefont {T.~A.}\ \bibnamefont
  {Loring}},\ }\href {https://arxiv.org/abs/1012.1019} {\bibfield  {journal}
  {\bibinfo  {journal} {Annals of Physics}\ }\textbf {\bibinfo {volume}
  {326}},\ \bibinfo {pages} {1699} (\bibinfo {year} {2011})}\BibitemShut
  {NoStop}%
\bibitem [{\citenamefont {Estienne}\ \emph {et~al.}(2012)\citenamefont
  {Estienne}, \citenamefont {Regnault},\ and\ \citenamefont
  {Bernevig}}]{Estienne2012}%
  \BibitemOpen
  \bibfield  {author} {\bibinfo {author} {\bibfnamefont {B.}~\bibnamefont
  {Estienne}}, \bibinfo {author} {\bibfnamefont {N.}~\bibnamefont {Regnault}},
  \ and\ \bibinfo {author} {\bibfnamefont {B.~A.}\ \bibnamefont {Bernevig}},\
  }\href {\doibase 10.1103/PhysRevB.86.241104} {\bibfield  {journal} {\bibinfo
  {journal} {Phys. Rev. B}\ }\textbf {\bibinfo {volume} {86}},\ \bibinfo
  {pages} {241104} (\bibinfo {year} {2012})}\BibitemShut {NoStop}%
\bibitem [{\citenamefont {Wang}\ \emph {et~al.}(2020)\citenamefont {Wang},
  \citenamefont {Rodriguez}, \citenamefont {Jiao}, \citenamefont {Howard},
  \citenamefont {Graham}, \citenamefont {Gu}, \citenamefont {Hughes},
  \citenamefont {Morr},\ and\ \citenamefont
  {Madhavan}}]{Vidyadensityinhomogeneity}%
  \BibitemOpen
  \bibfield  {author} {\bibinfo {author} {\bibfnamefont {Z.}~\bibnamefont
  {Wang}}, \bibinfo {author} {\bibfnamefont {J.~O.}\ \bibnamefont {Rodriguez}},
  \bibinfo {author} {\bibfnamefont {L.}~\bibnamefont {Jiao}}, \bibinfo {author}
  {\bibfnamefont {S.}~\bibnamefont {Howard}}, \bibinfo {author} {\bibfnamefont
  {M.}~\bibnamefont {Graham}}, \bibinfo {author} {\bibfnamefont {G.~D.}\
  \bibnamefont {Gu}}, \bibinfo {author} {\bibfnamefont {T.~L.}\ \bibnamefont
  {Hughes}}, \bibinfo {author} {\bibfnamefont {D.~K.}\ \bibnamefont {Morr}}, \
  and\ \bibinfo {author} {\bibfnamefont {V.}~\bibnamefont {Madhavan}},\ }\href
  {\doibase 10.1126/science.aaw8419} {\bibfield  {journal} {\bibinfo  {journal}
  {Science}\ }\textbf {\bibinfo {volume} {367}},\ \bibinfo {pages} {104}
  (\bibinfo {year} {2020})}\BibitemShut {NoStop}%
\bibitem [{\citenamefont {Bernevig}\ \emph {et~al.}(2006)\citenamefont
  {Bernevig}, \citenamefont {Hughes},\ and\ \citenamefont {Zhang}}]{BHZmodel}%
  \BibitemOpen
  \bibfield  {author} {\bibinfo {author} {\bibfnamefont {B.~A.}\ \bibnamefont
  {Bernevig}}, \bibinfo {author} {\bibfnamefont {T.~L.}\ \bibnamefont
  {Hughes}}, \ and\ \bibinfo {author} {\bibfnamefont {S.-C.}\ \bibnamefont
  {Zhang}},\ }\href {\doibase 10.1126/science.1133734} {\bibfield  {journal}
  {\bibinfo  {journal} {Science}\ }\textbf {\bibinfo {volume} {314}} (\bibinfo
  {year} {2006}),\ 10.1126/science.1133734}\BibitemShut {NoStop}%
\bibitem [{\citenamefont {Fu}\ and\ \citenamefont
  {Kane}(2007)}]{KaneZ2invariant}%
  \BibitemOpen
  \bibfield  {author} {\bibinfo {author} {\bibfnamefont {L.}~\bibnamefont
  {Fu}}\ and\ \bibinfo {author} {\bibfnamefont {C.~L.}\ \bibnamefont {Kane}},\
  }\href {\doibase 10.1103/PhysRevB.76.045302} {\bibfield  {journal} {\bibinfo
  {journal} {Physical Review B}\ }\textbf {\bibinfo {volume} {76}} (\bibinfo
  {year} {2007}),\ 10.1103/PhysRevB.76.045302}\BibitemShut {NoStop}%
\end{thebibliography}%
\end{document}